\documentclass[aps,nofootinbib,prd,preprint,preprintnumbers]{revtex4-1}
\usepackage{graphicx}
\usepackage{url}
\usepackage{xcolor}
\usepackage{amsmath}
\usepackage{amssymb}
\usepackage{hyperref}
\usepackage[normalem]{ulem}
\usepackage{soul}
\hypersetup{colorlinks=true,linkcolor=redLinks,citecolor=greenLinks,
urlcolor=redLinks,
pdfborder={0 0 1}}
\hypersetup{
    bookmarks=true,         
    unicode=false,          
    pdftoolbar=true,        
    pdfmenubar=true,        
    pdffitwindow=false,     
    pdfstartview={FitH},    
    pdftitle={My title},    
    pdfauthor={Author},     
    pdfsubject={Subject},   
    pdfcreator={Creator},   
    pdfproducer={Producer}, 
    pdfkeywords={keyword1, key2, key3}, 
    pdfnewwindow=true,      
    colorlinks=false,       
    linkcolor=red,          
    citecolor=green,        
    filecolor=magenta,      
    urlcolor=cyan           
}
\colorlet{shadecolor}{gray!15}
\definecolor{greenLinks}{rgb}{0, 0.6, 0} 
\definecolor{YELLOW}{rgb}{0.5, 0, 0.5} 
\definecolor{blueLinks}{rgb}{0, 0, 0.6}
\definecolor{redLinks}{rgb}{0.6, 0, 0}
\definecolor{tempText}{rgb}{0.55, 0.10,0.67}
\definecolor{eprintLinks}{rgb}{0.4, 0.4, 0.4}
\definecolor{journalLinks}{rgb}{0.6, 0, 0}

\usepackage[T1]{fontenc} 
\usepackage{hyperref}
\usepackage{float}

\newcommand {\black} {\color{black}}

\def\SM{$\mathrm{SU(3)_c \otimes SU(2)_L \otimes U(1)_Y}$ }
\def\21{$\mathrm{SU(2)_L \otimes U(1)_Y}$ }
\def\lfv{lepton flavour violation }
\def\cpv{CP violation }
\def\lnv{lepton number violation }

\newcommand{\sm}{standard model }

\newcommand{\AddrAHEP}{AHEP Group, Institut de F\'{i}sica Corpuscular --
  C.S.I.C./Universitat de Val\`{e}ncia, Parc Cientific de Paterna.\\
  C/Catedratico Jos\'e Beltr\'an, 2 E-46980 Paterna (Val\`{e}ncia) - SPAIN}

\newcommand{\MPIK}{Max-Planck-Institut f\"{u}r Kernphysik, Heidelberg 69117, Germany}

\definecolor{darkgreen}{RGB}{60,179,113}

\definecolor{gesfpurple}{rgb}{0.47,0.19,0.42}

\definecolor{gesflanse}{rgb}{0.00,0.50,0.50}

\definecolor{gesfblue}{rgb}{0.08,0.42,0.76}

\definecolor{gesfred}{rgb}{1,0,0}

\definecolor{gesfwhite}{rgb}{1,1,1}

\definecolor{gesfblack}{rgb}{0,0,0}

\def \znbb {$\rm 0\nu\beta\beta$ }
\newcommand{\gsec}[1]{{\hypersetup{linkcolor=red}Sec.~\ref{#1}\hypersetup{linkcolor=blue}}}
\newcommand{\gapp}[1]{{\hypersetup{linkcolor=red}App.~\ref{#1}\hypersetup{linkcolor=blue}}}
\newcommand{\geqn}[1]{\hypersetup{linkcolor=blue}(\ref{#1})\hypersetup{linkcolor=blue}}
\newcommand{\gfig}[1]{{\hypersetup{linkcolor=violet}Fig.~\ref{#1}\hypersetup{linkcolor=blue}}}
\newcommand{\gtab}[1]{{\hypersetup{linkcolor=gesflanse}Tab.~\ref{#1}\hypersetup{linkcolor=blue}}}

\def\31{$\mathrm{SU(3)_c \otimes U(1)_Q}$ }
\def\SM{$\mathrm{SU(3)_c \otimes SU(2)_L \otimes U(1)_Y}$ }

\newcommand{\beq}{\begin{equation}}
\newcommand{\eeq}{\end{equation}}                     
\newcommand{\beqa}{\begin{eqnarray}}
\newcommand{\eeqa}{\end{eqnarray}}          

\newcommand {\ignore}[1]{}

\graphicspath{{plots/}}

\begin{document}

\title{Measuring the Leptonic CP Phase in \\Neutrino Oscillations with Non-Unitary Mixing}

\author{Shao-Feng Ge~$^1$}\email{gesf02@gmail.com}
\author{Pedro Pasquini~$^{2,3}$}\email{pasquini@ifi.unicamp.br}
\author{M. T\'ortola~$^2$}\email{mariam@ific.uv.es}
\author{J. W. F. Valle~$^2$} \email{valle@ific.uv.es, URL:
  http://astroparticles.es/} 
\affiliation{$^1$~\MPIK}
\affiliation{$^2$~\AddrAHEP}
\affiliation{$^3$~Instituto de F\'isica Gleb Wataghin - UNICAMP, {13083-859}, Campinas SP, Brazil}

\begin{abstract}

  Non-unitary neutrino mixing implies an extra CP violating phase that
  can fake the leptonic Dirac CP phase $\delta_{CP}$ of the simplest
  three-neutrino mixing benchmark scheme. This would hinder the
  possibility of probing for CP violation in accelerator-type
  experiments. We take T2K and T2HK as examples to demonstrate the
  degeneracy between the ``standard'' (or ``unitary'') and
  ``non-unitary'' CP phases. We find, under the assumption of
  non-unitary mixing, that their CP sensitivities severely
  deteriorate. Fortunately, the TNT2K proposal of supplementing T2(H)K
  with a $\mu$DAR source for better measurement of $\delta_{CP}$ can
  partially break the CP degeneracy by probing both $\cos \delta_{CP}$
  and $\sin \delta_{CP}$ dependences in the wide spectrum of the
  $\mu$DAR flux. We also show that the further addition of a near
  detector to the $\mu$DAR setup can eliminate the degeneracy
  completely.

 \end{abstract}

 \pacs{13.15.+g,12.90.+b,23.40.Bw} 

 \maketitle

 \section{Introduction}

 The search for leptonic CP violation constitutes one of the major
 challenges in particle physics today~\cite{Branco:2011zb}. 
 Although CP violation studies are interesting in their own right, they
 may also shed light upon the general CP symmetries of the neutrino
 mass matrices in a rather model--independent way~\cite{Chen:2016ica},
 such as the case of the generalized $\mu-\tau$ reflection
 symmetry~\cite{Chen:2015siy}.
 Likewise, they can probe the predictions made by specific flavor
 models and hence put to test the structure of the corresponding
 symmetries~\cite{Morisi:2012fg,King:2014nza}.

 This type of CP violation is associated with the Dirac phase
 $\delta_{CP}$ present in the simplest three-neutrino mixing matrix,
 which is simply the leptonic analogue of the phase in the CKM matrix,
 describing the quark weak
 interactions~\cite{kobayashi:1973fv,Schechter:1980gr,PhysRevLett.51.1945}.
 It is known to directly affect lepton number conserving processes
 such as neutrino oscillations.
 So far neutrino oscillation experiments have measured the two squared
 neutrino mass differences, as well as the three corresponding mixing
 angles~\cite{Maltoni:2004ei}.  These measurements provide a rather
 precise determination of all neutrino oscillation parameters, except
 for the atmospheric mixing angle $\theta_{23}$, whose octant is still
 uncertain, and the leptonic Dirac CP phase $\delta_{CP}$, which is
 poorly determined~\cite{Forero:2014bxa}.
 The precision era in neutrino physics has come with new experimental
 setups that will provide enough statistics for measuring all of the
 neutrino parameters to an unprecedented level of accuracy. These
 include T2K~\cite{Abe:2015awa}, Hyper-K~\cite{Abe:2011ts}, and
 TNT2K~\cite{TNT2K}. The TNT2K (Tokai 'N Toyama to Kamioka) project is
 a combination of $\mu$Kam (with $\mu$DAR source and Super-K ($\mu$SK)
 or Hyper-K ($\mu$HK) detectors at Kamioka) and T2(H)K.

 All of the above facilities aim at measuring this single Dirac phase
 $\delta_{CP}$.
 However, one is likely to depart from such a simple picture, if
 neutrinos get their mass \textit{a la seesaw}.
 In this case, neutrino mass arises through the tree level exchange of
 heavy, so far undetected, \SM singlet messenger fermions such as
 ``right-handed'' neutrinos, as in the type-I seesaw mechanism.
 If the seesaw scheme responsible for generating neutrino mass is
 accessible to the LHC, then it is natural to expect that neutrino
 oscillations will be described by a non-unitary mixing matrix.
 Examples of such mechanisms are the inverse and linear seesaw
 schemes~\cite{Mohapatra:1986bd,GonzalezGarcia:1988rw,Akhmedov:1995vm,Akhmedov:1995ip,Malinsky:2005bi,Bazzocchi:2010dt}.
 In these schemes one expects sizeable deviations from the simplest
 three--neutrino benchmark, in which there are only three families of
 orthonormal neutrinos.

 The generic structure of the leptonic weak interaction was first given
 in Ref.~\cite{Schechter:1980gr} and contains new parameters in
 addition to those of the simplest three--neutrino paradigm. In this
 case the description of neutrino oscillations involves an effectively
 non-unitary mixing matrix~\cite{Escrihuela:2015wra,Li:2015oal}.
 As a consequence, there are degeneracies in the neutrino oscillation
 probability involving the ``standard'' three-neutrino CP phase and the
 ``new'' phase combination arising from the non-unitarity of the
 neutrino mixing matrix~\cite{Miranda:2016wdr,Miranda:2016ptb}.
 In this paper we examine some strategies to lift the degeneracies
 present between ``standard'' and ``new'' leptonic CP violation
 effects, so as to extract with precision the Dirac CP phase from 
 neutrino oscillations in the presence of non-unitary mixing. Such
 effort also provides an indirect way to help probing the mass scale
 involved in neutrino mass generation through the seesaw mechanism.
 A precise measurement of the genuine Dirac CP phase would also
 provide direct tests of residual symmetries that can predict
 correlation between the Dirac CP phase and the mixing angles
 \cite{Ge:2010js,He:2011kn,Dicus:2010yu,Ge:2011ih,Ge:2011qn,Hanlon:2013ska,He:2015xha}.

   Note also that probing the non-unitarity of the neutrino mixing
   matrix in oscillation searches could provide indirect indications
   for the associated (relatively low--mass) seesaw messenger
   responsible for inducing neutrino mass.
   This would also suggest that the corresponding charged \lfv and \cpv
   processes could be sizeable, irrespective of the observed smallness
   of neutrino
   masses~\cite{bernabeu:1987gr,branco:1989bn,rius:1989gk,Deppisch:2004fa,Deppisch:2005zm}.
   The spectrum of possibilities becomes even richer in low--scale
   seesaw theories beyond the \SM gauge
   structure~\cite{Deppisch:2013cya,Das:2012ii}.
 Unfortunately, however, no firm model--independent predictions can be
 made in the charged sector. As a result searches for the exotic
 features such as non--unitary neutrino propagation effects may provide
 a unique and irreplaceable probe of the theory that lies behind the
 canonical three--neutrino benchmark.

 This paper is organized as follows. In \gsec{sec:formalism} we
 summarize the generalized formalism describing neutrino mixing in the
 presence of non-unitarity. This convenient parametrization is then
 used to derive the non-unitarity effects upon the three--neutrino
 oscillation probabilities, by decomposing their dependence on the CP
 phases and the atmospheric mixing angle $\theta_a$, see details in
 \gapp{sec:decomposition}.  This is useful to demonstrate, in
 \gsec{sec:effect}, that the size of the non-unitary CP effects can be
 as large as the standard CP terms, given the current limits on
 leptonic unitarity violation. 
In addition, we also implement the
 inclusion of matter effects~\cite{Mikheev:1986gs,Wolfenstein:1977ue},
 as detailed in \gapp{sec:matter}, and illustrate how they can modify
 the oscillation probabilities.
 With the formalism established, we show explicitly in \gsec{sec:fake}
 how the ``non-unitary'' CP phase can fake the standard ``unitary'' one
 at accelerator neutrino experiments like T2(H)K.  In \gsec{sec:muDAR}
 we show that the degeneracy between unitary and non-unitary CP phases
 can be partially resolved with TNT2K. 
 Moreover, we further propose a near detector $\mu$Near, with 20\,ton
 of liquid scintillator and 20\,m of baseline, in order to disentangle
 the effects of the two physical CP phases and recover the full
 $\delta_{CP}$ sensitivity at TNT2K. Our numerical simulations for
 T2H(K), $\mu$SK, $\mu$HK, and $\mu$Near are carried out with the NuPro
 package \cite{NuPro}. The conclusion of this paper can be found in
 \gsec{sec:conclusion}. 

 \section{Neutrino Mixing Formalism}
 \label{sec:formalism}

 Within the standard three--neutrino benchmark scheme the neutrino
 flavor and mass eigenstates are connected by a unitary mixing matrix
 $U$~\cite{Valle:2015pba},
 \begin{equation}
   \nu_\alpha
 =
   U_{\alpha i} \nu_i \,,
         \label{eq:mix}
 \end{equation}
 where we use the subscript $\alpha$ for flavor and $i$ for mass
 eigenstates. This lepton mixing matrix may be expressed as
 \begin{equation}
   U
 =
   \mathcal P
 \left\lgroup
 \begin{array}{ccc} 
   c_s c_r & s_s c_r & s_r e^{- i \delta_{CP}} \\
 - c_a s_s - s_a s_r c_s e^{i \delta_{CP}} & c_a c_s - s_a s_r s_s e^{i \delta_{CP}} & s_a c_r \\
   s_a s_s - c_a s_r c_s e^{i \delta_{CP}} &-s_a c_s - c_a s_r s_s e^{i \delta_{CP}} & c_a c_r
 \end{array}
 \right\rgroup
   \mathcal Q \,.
 \label{eq:U}
 \end{equation}
 in which we have adopted the PDG variant~\cite{Agashe:2014kda} of the
 original symmetric parametrization of the neutrino mixing
 matrix~\cite{Schechter:1980gr}, with the three mixing angles
 $\theta_{12}$, $\theta_{23}$ and $\theta_{13}$ denoted as
 $\theta_{s}$, $\theta_{a}$ and $\theta_{r}$, for solar, atmospheric
 and reactor, respectively. 
 Within this description, three of the CP phases in the diagonal
 matrices $\mathcal P \equiv \mbox{diag}\{e^{- i \beta_1}, e^{- i
   \beta_2}, e^{- i \beta_3}\}$ and $\mathcal Q \equiv
 \mbox{diag}\{e^{- i \alpha_1}, e^{- i \alpha_2}, e^{- i \alpha_3}\}$
 can be eliminated by redefining the charged lepton fields, while one is
 an overall phase that can be rotated away. The remaining phases
 correspond to the two physical Majorana
 phases~\cite{Schechter:1980gr}~\footnote{The absence of invariance
   under rephasings of the Majorana neutrino Lagrangean leaves these
   extra two physical Majorana phases~\cite{Schechter:1980gr}. They do
   not affect oscillations~\cite{Schechter:1981gk,Doi:1980yb}, entering
   only in \lnv processes, such as neutrinoless double beta decay or
   \znbb \cite{Schechter:1981bd}.}.
 This leaves only the Dirac CP-phase $\delta_{CP}$ characterizing CP
 violation in neutrino oscillations.

 If neutrinos acquire mass from the general seesaw mechanism through
 the exchange of \SM singlet heavy messenger fermions, these extra
 neutrino states mix with the standard $\nu_e$, $\nu_\mu$, 
 $\nu_\tau$, and then the neutrino mixing needs to be extended to go beyond $3
 \times 3$,
 \begin{equation}
   U^{n \times n}
 =
 \left\lgroup
 \begin{matrix}
   N & W \\
   V & T
 \end{matrix}
 \right\rgroup \,,
 \label{eq:Unn}
 \end{equation}
 Note that the total mixing matrix $U^{n \times n}$ (with $n > 3$)
 shall always be unitary, regardless of its size.  The leptonic weak
 interaction mixing matrix is promoted to rectangular
 form~\cite{Schechter:1980gr} where each block can be systematically
 determined within the seesaw expansion~\cite{Schechter:1981cv}.
 However if the extra neutrinos are heavy they cannot be produced at
 low energy experiments nor will be accessible to oscillations.
 In such case only the first $3 \times 3$ block $N$ can be
 visible~\cite{valle:1987gv,nunokawa:1996tg,Antusch:2006vwa}. In other
 words, the original $3 \times 3$ unitary mixing $U$ in \geqn{eq:U} is
 replaced by a truncated non-unitary mixing matrix $N$ which will
 effectively describe neutrino propagation.  This can be written as
 \begin{equation}
   N
 =
   N^{NP} U
 =
 \left\lgroup
 \begin{array}{ccc} 
 \alpha_{11} & 0 & 0\\
 \alpha_{21} & \alpha_{22} & 0\\
 \alpha_{31} & \alpha_{32} & \alpha_{33}
 \end{array}
 \right\rgroup U \,.
 \label{eq:N}
 \end{equation}
 This convenient parametrization follows from the symmetric one in
 \cite{Schechter:1980gr} and applies for any number of additional
 neutrino states~\cite{Escrihuela:2015wra}. Irrespective of
   the number of heavy singlet neutrinos, it involves three real
 parameters ($\alpha_{11},\alpha_{22}$ and $\alpha_{33}$, all close to
 one) and three small complex parameters ($\alpha_{21},\alpha_{31}$ and
 $\alpha_{32}$). In the \sm one has, of course, $\alpha_{ii}=1$ and
 $\alpha_{ij}=0$ for $i\neq j$.
  Current experiments, mainly involving electron and muon
   neutrinos, are sensitive to three of these parameters:
   $\alpha_{11}$, $\alpha_{22}$ and $\alpha_{21}$. Note that the latter
   is complex and therefore we end up with three additional real
   parameters and one new complex phase $$\phi \equiv
   -{\rm arg}(\alpha_{21}).$$
 The above definition matches the notation in
   Refs.~\cite{Escrihuela:2015wra,Miranda:2016wdr}.

   There are a number of constraints on non-unitarity, such as those
   that follow from weak universality
   considerations. In~\cite{Escrihuela:2015wra} updated constraints on
   unitarity violation parameters at 90\% C.L. have been given as
 \begin{equation}
   \alpha_{11}^2 \geq 0.989 \,,
 \quad 
   \alpha_{22}^2 \geq 0.999 \,,
 \quad 
   |\alpha_{21}|^2 \leq 6.6 \times 10^{-4} \,,
   \label{eq:bounds}
 \end{equation}
 These include both universality as well as oscillation limits.
 Concerning the former, these constraints are all derived on the basis
 of charged current induced processes and under the assumption that
 there is no new physics other than that of non-unitary mixing.
 Such bounds rely on many simplifying assumptions. Departure from such
 simplifying approximations could result in different bounds on the
 non-unitarity parameters.

 Indeed, although naively one might think that new physics
 interactions would always enhance the deviation from the \sm
 prediction, strengthening the non-unitarity bounds, the opposite can
 happen.
 For example, new physics can weaken the non-universality bounds as a
 result of subtle cancellations involving the new physics effects
 contributing to the relevant weak processes~\footnote{ Though
   less likely, cancellations between new physics and \sm
   contributions to a given weak process can also be envisaged.  }. It
 is not inconceivable that such cancellations amongst new physics
 contributions might even result from adequately chosen symmetry
 properties of the new interactions.  


 Given the fragility of existing constraints, the main emphasis of our
 paper will be on experiments providing robust model-independent
 bounds on non-unitarity relying only on neutrino processes.
 For this reason here we will concentrate on the following bound on
 $\alpha_{21}$ due the non-observation of $\nu_\mu$ to $\nu_e$
 conversion at the NOMAD experiment, only relevant neutrino
 oscillation experiment. We implement this bound as \textit{prior} in
 the NuPro package~\cite{NuPro} as
 \begin{equation}
     \left[ \sin^2 (2 \theta_{\mu e}) \right]_{eff}
 = 
   2 |\alpha_{21}|^2
 \leq
   0.0014
\quad \quad @ \,\, 90\% \text{C.L.}\,
 \label{eq:prior2}
  \end{equation}
 %

  In contrast to non-oscillation phenomena, the NOMAD experiment puts
  direct constraints on neutrino oscillations, which can be used as a
  prior in our simulation.  Indeed, the presence of new physics
  affecting the charged lepton sector would not change the previous
  bound, since NOMAD results were derived by assuming the standard
  model values for observables such as $R^\pi_{e\mu}$. These values
  are in agreement with current experimental observations and
  therefore they will not be affected by any other process of new
  physics in the charged sector.
    In contrast, new physics in the neutrino sector such as
    non-standard interactions with matter or light sterile neutrinos
    could affect the bound in Eq.~(\ref{eq:prior2}). Besides, these
    additional physics phenomena would have in general different
    effects in NOMAD and T2K and therefore the above limit will not be
    directly applicable to T2K. In order to simplify the physics
    scenario, here we focus on non-unitarity as the only source of new
    physics in the neutrino sector.  
  Since no sensitivity on the non-unitary CP phase $\phi$ has been
  obtained so far so we will take this parameter free in our analysis.
  We will show how non-unitary mixing can deteriorate the CP
  measurement in neutrino oscillation experiments under the current
  model-independent constraints. What we propose in this paper can
  improve not only the constraint on non-unitary mixing but also the
  resulting CP sensitivity~\cite{Miranda:2016wdr}.
  As a reference benchmark value for $\alpha_{21}$ we may take the
  above bound given by the NOMAD experiment.

 \section{Effect of The Non-Unitarity CP Phase}
 \label{sec:effect}

 As demonstrated in \cite{Ge:2013zua}, the three currently unknown
 parameters in neutrino oscillations, the neutrino mass hierarchy, the
 leptonic Dirac CP phase $\delta_{CP}$, and the octant of the
 atmospheric angle $\theta_a$, can be analytically disentangled from
 each other.  This decomposition formalism is extremely useful to study
 the effect of different unknown parameters in various types of
 neutrino oscillation experiments. Here, we generalize the formalism to
 accommodate the effect of non-unitary neutrino mixing, $N = N^{NP} U$,
 as parametrized in Eq.~\geqn{eq:N}. This extra mixing can be
 factorized from the Hamiltonian $\mathcal H^{NP}$ and the oscillation amplitude $S^{NP}$, 
 together with $U_{23}(\theta_a)$, which is the 2--3 mixing due to the atmospheric angle $\theta_a$, and
 the rephasing matrix $P_\delta \equiv \mbox{diag}(1,1,e^{i
   \delta_{CP}})$,
 \begin{subequations}
 \begin{eqnarray}
   \mathcal H^{NP}
 & = &
   [N^{NP} U_{23}(\theta_a) P_\delta] \mathcal H' [N^{NP} U_{23}(\theta_a) P_\delta]^\dagger \,,
 \\
   S^{NP}
 & = &
   [N^{NP} U_{23}(\theta_a) P_\delta] S' [N^{NP} U_{23}(\theta_a) P_\delta]^\dagger \,.
 \end{eqnarray}
 \end{subequations}
 With less mixing parameters, it is much easier to first evaluate $S'$
 with the transformed Hamiltonian $\mathcal H'$. The effect of the
 non-unitary mixing parameters in $N^{NP}$, the atmospheric angle
 $\theta_a$ and the Dirac CP phase $\delta_{CP}$ can then be retrieved
 in an analytical way (see \gapp{sec:decomposition} for more details).

 Here, we find that the key oscillation probability $P_{\mu e}$ for the
 $\nu_\mu \to \nu_e$ channel is given by,
 \begin{eqnarray}
   P^{NP}_{\mu e}
 & = &
   \alpha^2_{11}
 \left\{
   \alpha^2_{22}
 \left[
   c^2_a |S'_{12}|^2
 + s^2_a |S'_{13}|^2
 + 2 c_a s_a (\cos \delta_{CP} \mathbb R - \sin \delta_{CP} \mathbb I) (S'_{12} S'^*_{13})
 \right]
 + |\alpha_{21}|^2 P_{ee}
 \right.
 \nonumber
 \\
 &+&
 \left.
   2 \alpha_{22} |\alpha_{21}|
 \left[
   c_a \left( c_\phi \mathbb R - s_\phi \mathbb I \right) (S'_{11} S'^*_{12})
 + s_a \left( c_{\phi + \delta_{CP}} \mathbb R - s_{\phi + \delta_{CP}} \mathbb I \right) (S'_{11} S'^*_{13})
 \right]
 \right\} \,.
 \label{eq:PmeNP}
 \end{eqnarray}
 The choice of this parametrization is extremely convenient to separate
 the neutrino oscillation probabilities into several terms, as we
 further elaborate in \gapp{sec:decomposition}.
 In this formalism, the transition probability $P^{NP}_{\mu e}$
 relevant for the CP studies can be decomposed into several terms,
 $P^{NP}_{\mu e} = \sum_k f_k(\alpha_{ij}, \theta_a, \phi) P^{(k)}_{\mu
   e}(S')$.  It contains six terms $P^{(2,3,7,8,9,10)}_{\mu e}$
 involving the Dirac CP phases $\delta_{CP}$ and $\phi$ (see Table
 \ref{tab:Ps} in \gapp{sec:decomposition}). The standard phase
 $\delta_{CP}$ is modulated by $P^{(2,3)}_{\mu e}$, which are mainly
 controlled by the matrix elements $(\mathbb R, \mathbb I)(S'_{12}
 S'^*_{13})$, while the non-unitarity counterparts $P^{(7,8,9,10)}_{\mu
   e}$ involve the elements $(\mathbb R, \mathbb I)(S'_{11} S'^*_{12},
 S'_{11} S'^*_{13})$.

\begin{figure}[t]
 \centering
 \includegraphics[width=8cm]{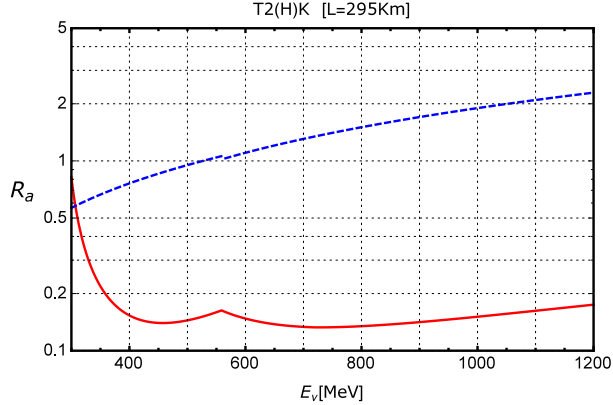}
 \caption{$R_a$ ratio as given in Eq. (\ref{eq:Ratios}) for the
   T2(H)K experimental setup, setting
   $2|\alpha_{21}|/\alpha_{22}=5\%$. 
 The solid red line corresponds to $R_2$, while $R_3$ is given by the dashed blue line.}
 \label{fig:Ratio-T2K}
 \end{figure}

 If $(\mathbb R, \mathbb I)(S'_{11} S'^*_{12}, S'_{11} S'^*_{13})$ are of
 the same size as $(\mathbb R, \mathbb I)(S'_{12} S'^*_{13})$, the
 effect of the non-unitary CP phase $\phi$ is then suppressed by the
 constraint $|\alpha_{21}| \lesssim 0.026$.
 Nevertheless, $S'_{11}$ has much larger magnitude than $S'_{12}$ and
 $S'_{13}$ which becomes evident by calculating the amplitude matrix
 $S'$ in the basis in which the atmospheric angle $\theta_a$ and the
 Dirac CP phase are factorized. Since the matter effects are small for
 the experiments under consideration, here we can illustrate the
 picture with the result in vacuum~\footnote{Although our results are
   obtained under the assumption that there is no matter effect, they
   also apply when the matter effect is not significant. 
   See \gapp{sec:matter} for details.},
 \begin{equation}
   S'
 =
   \mathbb I_{3 \times 3}
 -
   2 i \sin \Phi_a e^{- i \Phi_a}
 \left\lgroup
 \begin{matrix}
   s^2_r & & c_r s_r \\
 & 0 \\
   c_r s_r & & c^2_r
 \end{matrix}
 \right\rgroup
 -
   2 i \sin \Phi_s e^{- i \Phi_s}
 \left\lgroup
 \begin{matrix}
   c^2_r s^2_s        &  c_r c_s s_s &-c_r s_r s^2_s \\
   c_r c_s s_s        &  c^2_s       &-s_r c_s s_s   \\
 - c_r s_r s^2_s      & -s_r c_s s_s & s^2_r s^2_s        
 \end{matrix}
 \right\rgroup ,
 \label{eq:S'}
 \end{equation}
 where $\mathbb I_{3 \times 3}$ is the $3 \times 3$ identity matrix and
 $\Phi_{a,s} \equiv \Delta m^2_{a,s} / 4 E_\nu$ denote the solar and
 atmospheric oscillation phases. One can see explicitly that the
 amplitude matrix $S'$ is symmetric in the absence of matter potential
 as well as for symmetric matter profiles.
 \begin{figure}[t]
 \centering
 \includegraphics[width=8cm,angle=-90]{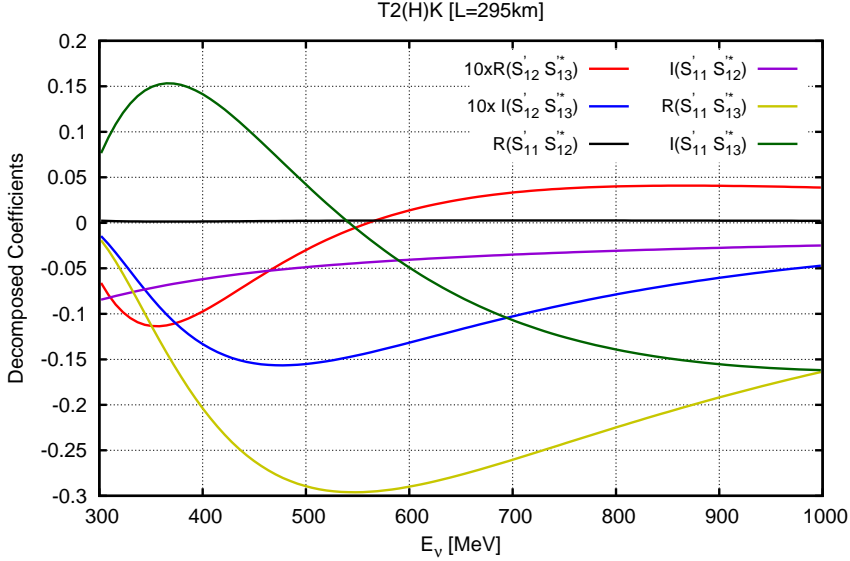}
 \caption{The decomposed CP coefficients for the neutrino 
          oscillation probability $P_{\mu e}$ for T2(H)K.}
 \label{fig:deCoeff-T2K}
 \end{figure}
 For CP measurements at accelerator experiments, the neutrino energy
 and baseline are usually configured around the first oscillation peak,
 $\Phi_a \approx \frac \pi 2$. Correspondingly, $\Phi_s \approx \frac
 \pi 2 \times \Delta m^2_s / \Delta m^2_a$, has a small value.  Up to
 leading order, $S'_{11} \approx 1$, in comparison with $S'_{12}
 \approx - 2 i \sin \Phi_s e^{- i \Phi_s} c_r c_s s_s$ and $S'_{13}
 \approx - 2 i \sin \Phi_a e^{- i \Phi_a} c_r s_r$.  The $S'_{12}$
 element is suppressed by $\Delta m^2_s / \Delta m^2_a$ while $S'_{13}$
 is suppressed by the reactor angle $\theta_r$. Consequently, the
 non-unitary elements $\mathbb I(S'_{11} S'^*_{12})$ and $(\mathbb R,
 \mathbb I)(S'_{11} S'^*_{13})$ are expected to be at least one order
 of magnitude larger than the unitary elements $(\mathbb R, \mathbb
 I)(S'_{12} S'^*_{13})$. Note that $S'_{12}$ is mainly imaginary, which
 makes $\mathbb R(S'_{11} S'^*_{12})$ to almost vanish.  Among the
 remaining non-unitary terms, there is still a hierarchical
 structure. Since $S'_{12}$ is suppressed by $\Delta m^2_s/\Delta
 m^2_a$ while $S'_{13}$ is suppressed by $s_r$, the relative size is
 roughly $|S'_{12} / S'_{13}| \sim 1/5$.  In short, there are five
 independent CP terms in $P_{\mu e}$, in full agreement with the result
 in \cite{Escrihuela:2015wra}.  To give an intuitive picture, we plot
 in \gfig{fig:deCoeff-T2K} the six CP related decomposition
 coefficients at T2(H)K \cite{TNT2K} for illustration.
 The relative size of the coefficients can then be measured by,
 \begin{equation}
 R_{a}\equiv\frac{2|\alpha_{21}|}{\alpha_{22}}\frac{\mathbb R (S'_{11}S'^*_{1a})+\mathbb I (S'_{11}S'^*_{1a})}{\mathbb R (S'_{12}S'^*_{13})+\mathbb I (S'_{12}S'^*_{13})},
 \label{eq:Ratios}
 \end{equation}
 where $a=2,3$.  We plot the ratio $R_{a}$ for
 $2|\alpha_{21}|/\alpha_{22}^2=5\%$ on \gfig{fig:Ratio-T2K}, where it
 is even clearer that $\mathbb I(S'_{11} S'^*_{12})$ and $(\mathbb R,
 \mathbb I)(S'_{11} S'^*_{13})$ are typically $\sim$ 10-20 times larger
 than $(\mathbb R, \mathbb I)(S'_{12} S'^*_{13})$, as expected.  These
 considerations show that the size of the standard and the non-unitary
 contribution can be of the same order. As a result, it can easily
 mimic the shape of the oscillation curve visible to the experimental
 setup.
 \begin{figure}[thb]
 \centering
 \includegraphics[scale=0.6]{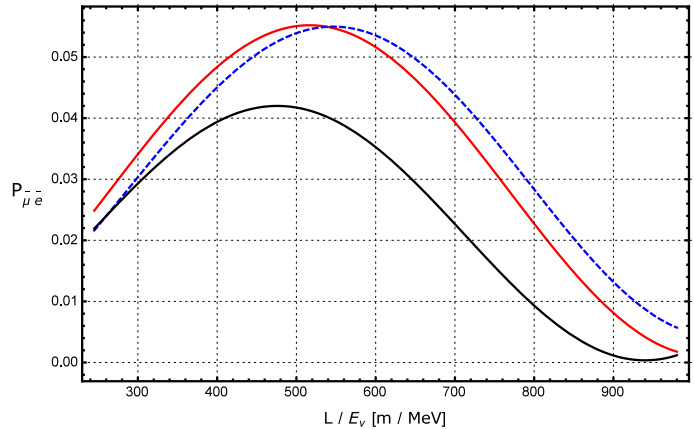} 
 \caption{Electron antineutrino appearance probability as a function of
   $L/E$ for three different assumptions: (i) black solid line: unitary
   case with $\delta_{CP}=0$, (ii) blue dashed line: unitary with
   $\delta_{CP}=3\pi/2$, (iii) red solid line: non-unitary case with
   $\delta_{CP}=0$, $|\alpha_{21}|=0.02$ and
   $\phi=0.1\pi$. 
 }
 \label{fig:mimic}
 \end{figure}

 Another intuitive way to observe this is through the plot of
 oscillation probability as a function of $L/E$ as in
 \gfig{fig:mimic}. Notice how a non-zero value of $\phi$ can mimic the
 behaviour of $\delta_{CP}=3\pi/2$ (dashed blue line) even with
 $\delta_{CP}=0$ (solid red line).  Later on, it will become clear that
 if the magnitude of the non-unitarity CP effect $|\alpha_{21}|$ is as
 large as $5 \%$, the standard CP phase $\delta_{CP}$ will not be
 distinguishable from its non-unitary counterpart $\phi$, unless the
 experiment can measure neutrino oscillations over a wide range of
 $L/E$. This issue will be taken up and elaborated in \gsec{sec:fake}.

\begin{figure}[t]
 \centering
 \includegraphics[scale=0.6]{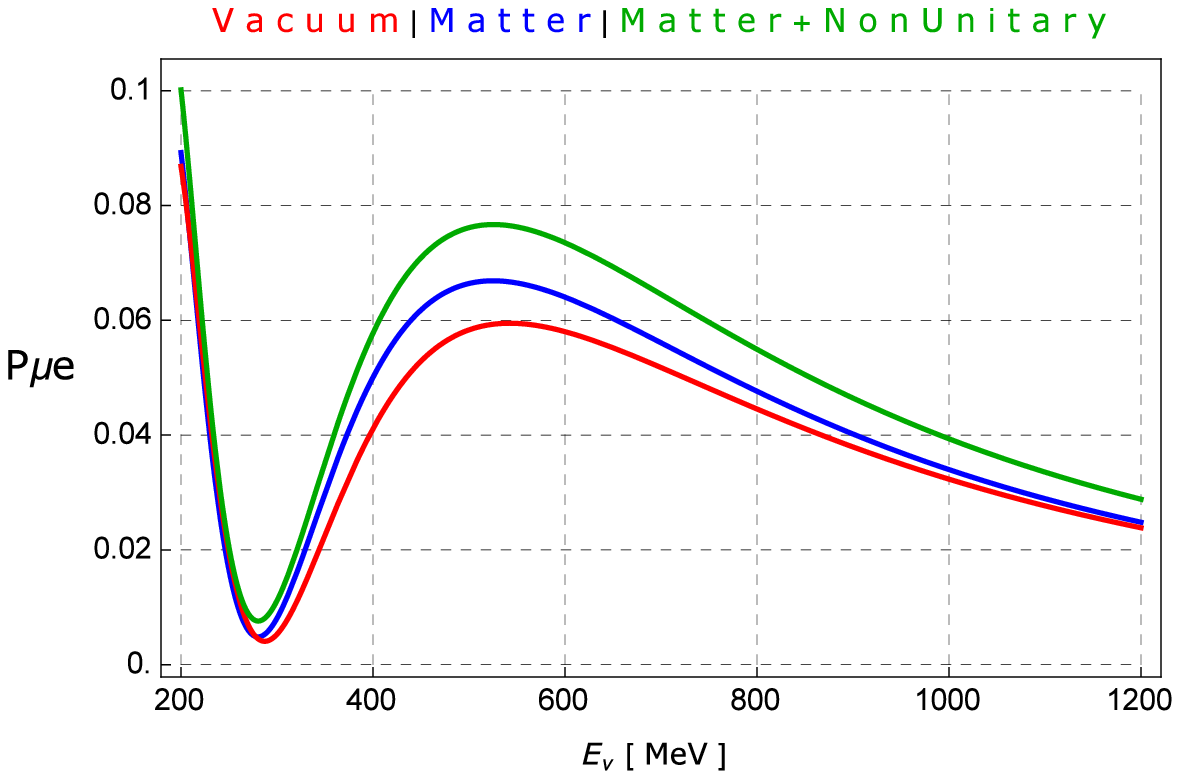}
  \includegraphics[scale=0.6]{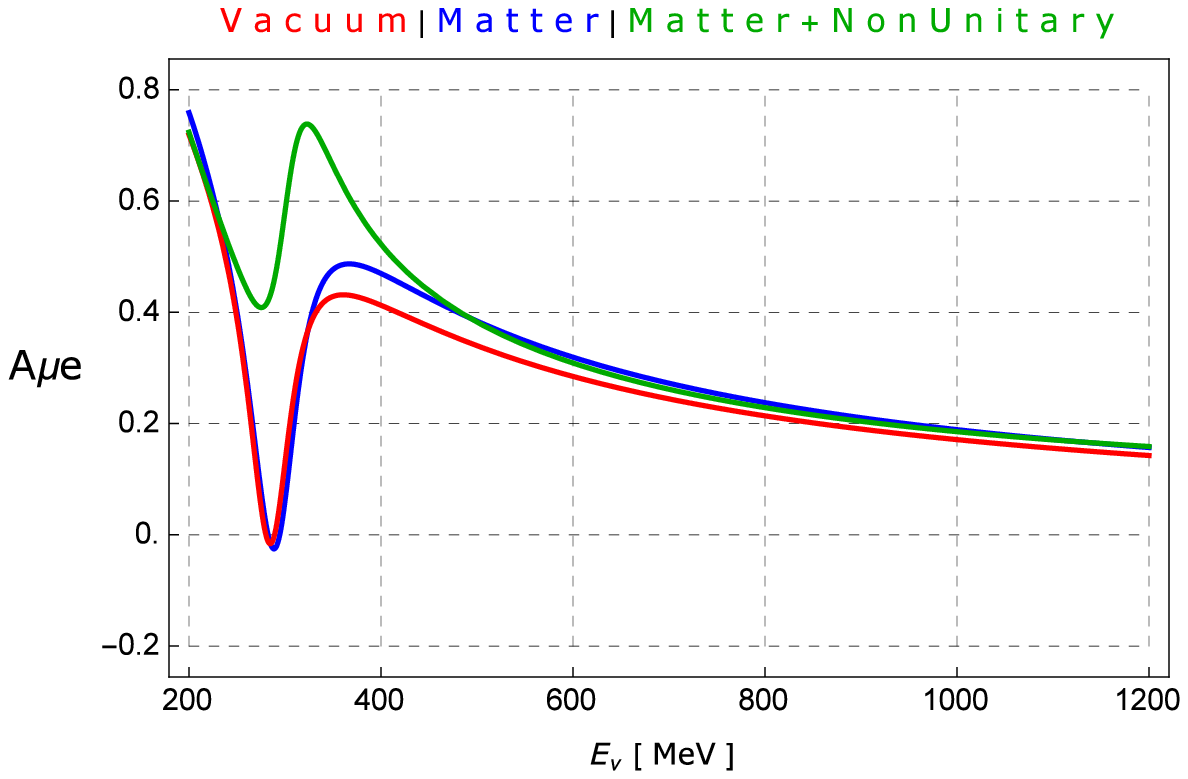}
  \caption{Left: muon to electron neutrino appearance
    probability at a baseline of $295$ km. Right: the corresponding CP
    asymmetry between neutrino and anti-neutrino oscillations. We
    compare three assumptions: unitary mixing in vacuum (red), unitary
    mixing in matter (blue) and non-unitary mixing in matter with
    $|\alpha_{21}|=0.02$ and $\phi=3\pi/2$ (green). In all cases we
    take $\delta_{CP}=3\pi/2$. }
 \label{fig:non-uni-matter}
 \end{figure}

 It should be pointed out that although in the T2K experiment the
 matter effect is small, it is not completely negligible when
 considering the sensitivity on the CP phases. The effect of the non-unitary mixing
 and the matter potential in the electron neutrino appearance probability is shown in 
 \gfig{fig:non-uni-matter}.
 This means
 that a CP analysis should take matter effects into account: in
 \gapp{sec:matter} we present a formalism to deal with matter effects
 in the context of non-unitary neutrino mixing. As a good
 approximation, one can assume an Earth profile with constant density
 $\rho_{\rm earth}=3 \,\rm{g}/\rm{cm}^3$ throughout this
 paper.

 \section{Faking the Dirac CP Phase with Non-Unitarity}
 \label{sec:fake}

 As depicted in Figs.~\ref{fig:deCoeff-T2K} and \ref{fig:Ratio-T2K},
 the size of the amplitude matrix elements
 $\mathbb I(S'_{11} S'^*_{12})$ and
 $(\mathbb R, \mathbb I)(S'_{11} S'^*_{13})$ that contribute to the CP
 terms associated to unitarity violation are typically $\sim$ 10-20
 times larger than their unitary counterparts
 $(\mathbb R, \mathbb I)(S'_{12} S'^*_{13})$.  According to the prior
 constraint in Eq.\geqn{eq:prior2}, the magnitude of the non-unitary
 CP term $|\alpha_{21}|$ is about $2.6\%$ at 90\% C.L. Consequently,
 after taking into account the extra factor of $2$ associated with
 $|\alpha_{21}|$ in \gtab{tab:Ps}, one finds that the non-unitary CP
 coefficients $P^{(8,9,10)}_{\mu e}$ can be as large as the unitary
 ones $P^{(2,3)}_{\mu e}$. Hence there is no difficulty for the
 non-unitary CP phase $\phi$ to fake the effects normally ascribed to
 the conventional CP phase $\delta_{CP}$, given the currently
 available prior constraint on non-unitarity.

\begin{figure}[t!]
 \centering
 \includegraphics[scale=0.31,angle=-90]{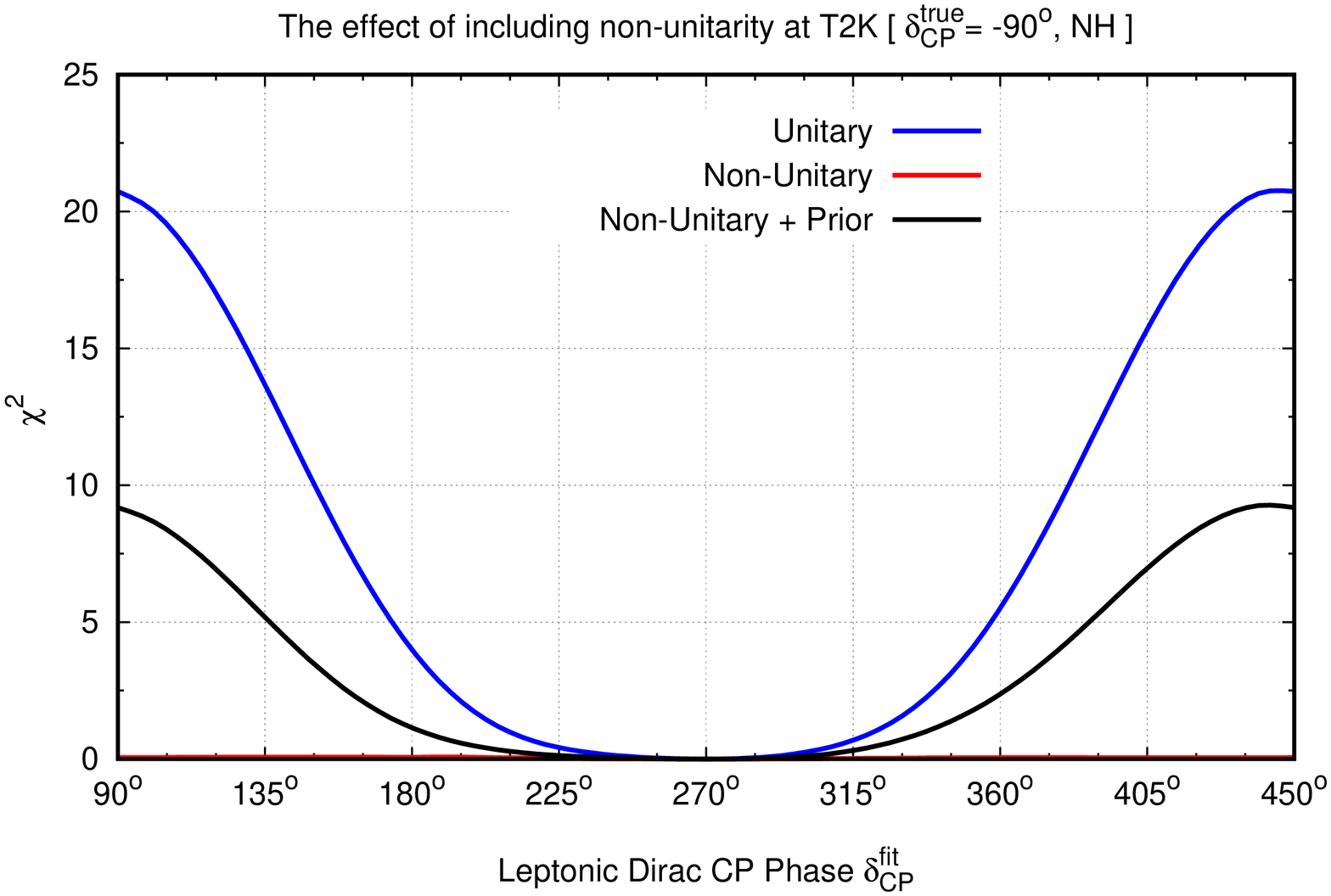}
 \includegraphics[scale=0.31,angle=-90]{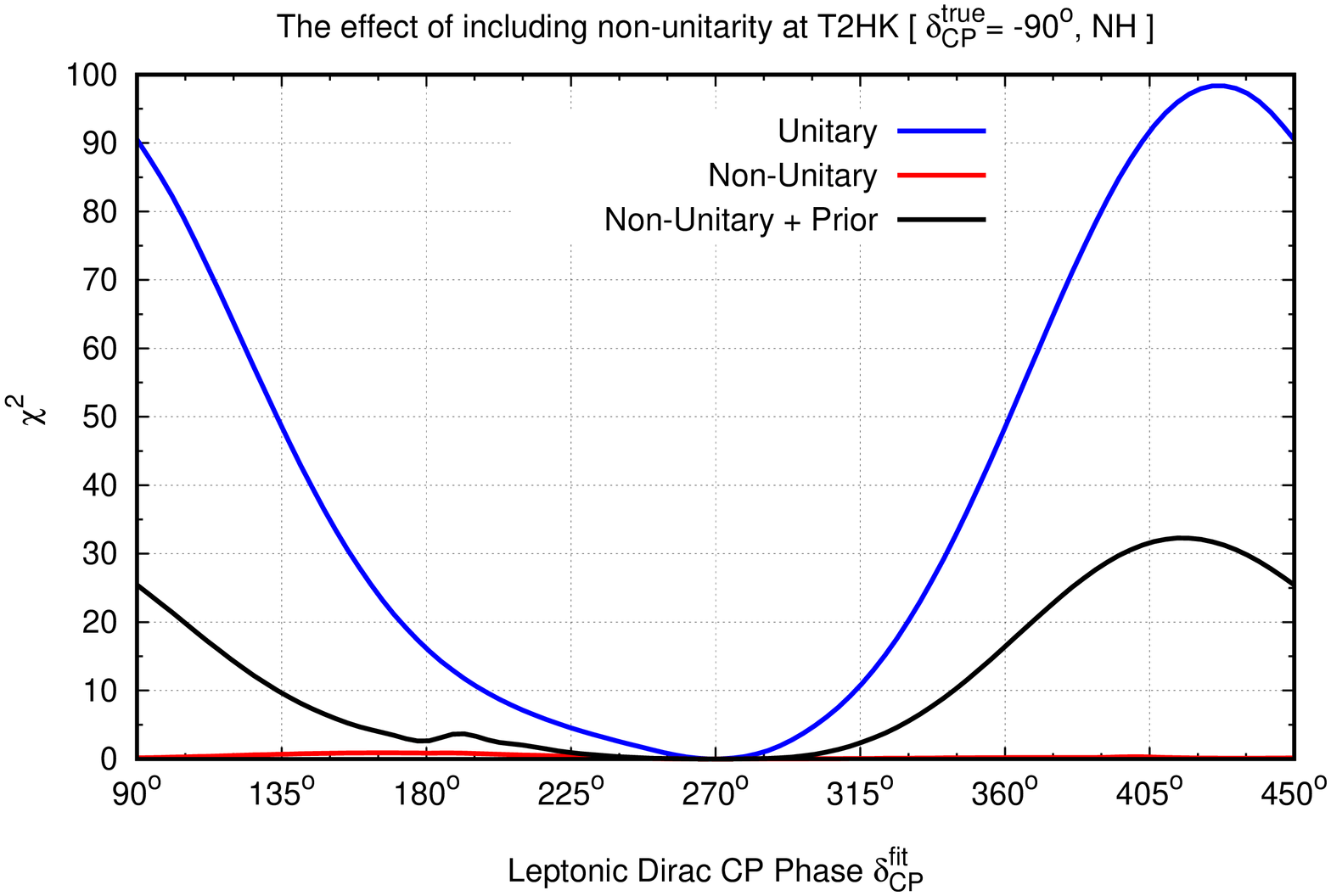}
 \caption{The marginalized $\chi^2(\delta_{CP})$ function at T2K and
   T2HK under the assumptions of unitary mixing ({\color{blue}{blue}}) and
   non-unitary mixing with ({\color{red}{red}}) or without (black)
   the prior constraint.}
 \label{fig:chi2-T2K}
 \end{figure}

 In order to study to what extent the standard CP phase $\delta_{CP}$
 can be faked by the non-unitary CP phase $\phi$, we simulate, for
 illustration, the T2(H)K experiment, as shown in \gfig{fig:chi2-T2K}.
 The pseudo-data are simulated with the true value of
 $\delta_{CP}=3\pi/2$, under the assumption of unitary mixing,
 \begin{equation}
   \delta^{true}_{CP} = 3\pi/2 \,,
 \qquad
   \alpha^{true}_{11} = \alpha^{true}_{22} = 1 \,,
 \qquad
   |\alpha_{21}|^{true} = 0 \,.
 \label{eq:true}
 \end{equation}
 In other words, there is no unitarity violation in the simulated
 pseudo-data.  We assume that the $7.8 \times 10^{21}\mbox{POT}$
   flux of T2K \cite{T2K1409}, corresponding to 6 years of running, is
   equally split between the neutrino and anti-neutrino modes, while
   the same configuration is assigned for T2HK in this section.

 To extract the sensitivity on the leptonic Dirac CP phase
 $\delta_{CP}$, we fit the pseudo-data with the following $\chi^2$
 function,
 \begin{equation}
   \chi^2
 \equiv
   \chi^2_{stat}
 + \chi^2_{sys}
 + \chi^2_{prior} \,,
 \end{equation}
 where the three terms ($\chi^2_{stat}$, $\chi^2_{sys}$,
 $\chi^2_{prior}$) stand for the statistical, systematical, and prior
 contributions. The statistical contribution $\chi^2_{stat}$ comes from
 the experimental data points,
 \begin{equation}
   \chi_{stat}^2
 =
   \sum_i
 \left(
   \frac{N_i^{\rm pred} - N_i^{\rm data}}{\sqrt{N_i^{\rm data}}}
 \right)^2 \,,
 \label{eq:chi2-stat}
 \end{equation}
 with summation over energy bins, for a specific experiment. For the
 combined analysis of several experiments, the total $\chi^2_{stat}$
 will be a summation over their contributions.  In the systematical
 term $\chi^2_{sys}$ we take into account the flux uncertainties.  For
 T2(H)K, we assume a 5\% flux uncertainty for the neutrino and
 anti-neutrino modes independently,
 \begin{equation}
   \chi^2_{sys}
 =
   \left( \frac {f_\nu - 1}{0.05} \right)^2
 + \left( \frac {f_{\bar \nu} - 1}{0.05} \right)^2 \,.
 \end{equation}
 Note that both the statistical $\chi^2_{stat}$ and systematical
 $\chi^2_{sys}$ parts need to be extended when adding extra
 experiments. In contrast, the prior knowledge is common for different
 experimental setups.  For the discussion that follows, it consists of
 two parts,
 \begin{equation}
   \chi^2_{prior}
 =
   \chi^2_{unitary}
 + \chi^2_{non-unitary} \,.
 \end{equation}
 The first term $\chi^2_{unitary}$ contains the current measurement of
 the three-neutrino oscillation parameters~\cite{Forero:2014bxa}, as
 summarized in the Sec.2.1 of \cite{TNT2K}, while the contribution
 $\chi^2_{non-unitary}$ accounts for the current constraint on the
 unitarity violating parameters in Eq.~\geqn{eq:prior2}. Note
 that the unitary prior contribution $\chi^2_{unitary}$ is always
 imposed while $\chi^2_{non-unitary}$ is only considered when fitting
 the data under the non-unitarity assumption with prior constraint.

 We then fit the data under different assumptions. For each value of
 the CP phase $\delta_{CP}$, the marginalized value of $\chi^2$ in
 \gfig{fig:chi2-T2K} is obtained by first fixing the fit value of
 $\delta_{CP}$ and then minimizing the $\chi^2$ function over the other
 oscillation parameters.  Depending on the assumption, the parameter
 list includes the three mixing angles, the two mass squared
 differences, and the non-unitary parameters.  The blue curves in
 \gfig{fig:chi2-T2K} are obtained by assuming standard unitary mixing,
 with minimization over the three mixing angles ($\theta_a$,
 $\theta_r$, $\theta_s$) and the two mass splittings ($\Delta m^2_a$,
 $\Delta m^2_s$).
 The result is the marginalized $\chi^2 (\delta_{CP})$ function from
 which we can read off the CP measurement sensitivity,
 $\chi^2(\delta_{CP}) = 1$ for $1 \sigma$. One can see that T2K can
 distinguish reasonably well a nonzero Dirac CP phase from zero, while
 T2HK can further enhance this sensitivity, under the unitarity
 assumption.
 We then turn on the non-unitarity parameters and
 $\chi^2_{non-unitary}$.  As we can see, the situation totally
 changes once non-unitarity is introduced. The inclusion of the
 non-unitarity degrees of freedom ($\alpha_{11}$, $\alpha_{22}$,
 $|\alpha_{21}|$, and $\phi$) requires the marginalization over nine
 parameters.
 Given a nonzero fitting value $\delta^{fit}_{CP}$, one can find a
 counter-term from the non-unitarity terms $P^{(8,9,10)}_{\mu e}$ that
 cancel the CP effect arising from the standard terms $P^{(2,3)}_{\mu
   e}$, leading to better agreement with the pseudo-data.
 In other words, the effect of the CP phase $\delta_{CP}$ can be faked
 by its non-unitary counterpart $\phi$. The resulting
 $\chi^2(\delta_{CP})$ becomes nearly flat, as shown by the red curves
 in \gfig{fig:chi2-T2K}. Under the assumption of non-unitary mixing,
 there is almost no CP sensitivity in either T2K or T2HK. 
 \begin{figure}[t]
 \centering
 \includegraphics[scale=0.6]{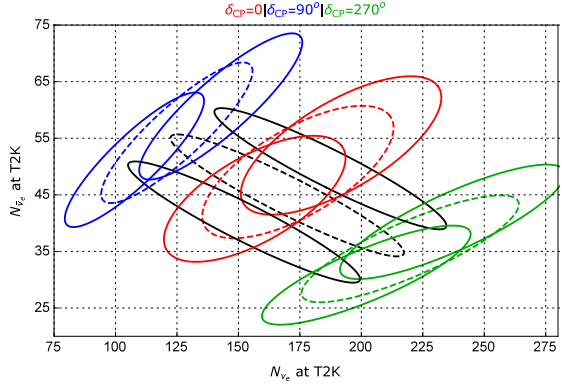}
 \caption{ Bi-event rate plot for T2K for standard three--neutrino
   mixing with varying $\delta_{CP}$ (black line), and non-unitary
   mixing with fixed $\delta_{CP}$ value and varying $\phi$ (color
   lines).  Dashed lines correspond to $\sin^2\theta_{a}=0.5$ while
   solid lines correspond to $\sin^2\theta_{a}=0.5\pm0.055$.  }
 \label{fig:elpseT2K}
 \end{figure}
 Imposing the correlated prior constraint \geqn{eq:prior2} as
 $\chi^2_{non-unitary}$ slightly improve the situation, shown as the
 black curves in \gfig{fig:chi2-T2K}.  Nevertheless, the CP
 sensitivity is still much worse than the standard case. The
 difference between $\delta^{true}_D = -90^o$ and
 $\delta^{fit}_D = 180^o$ reduces from $2\sigma$ to less than
 $1\sigma$.  With or without the prior constraint, the CP sensitivity at
 T2(H)K is significantly reduced by the presence of non-unitary
 mixing.

 An intuitive plot to illustrate this fact is presented in
 \gfig{fig:elpseT2K} where we show the event rates for the neutrino and
 antineutrino appearance channel in T2K for two different assumptions:
 the standard three--neutrino case with varying $\delta_{CP}$ (black
 line), and the alternative non-unitary case with fixed $\delta_{CP}$
 and varying $\phi$ (color lines).
 The variation of the atmospheric angle $\theta_a$ has been also considered in the non-unitary case.
 In particular, dashed lines in the plot correspond to maximal mixing, $\sin^2\theta_{a}=0.5$, while
   solid lines  cover approximately the 1$\sigma$ allowed range,
 $\sin^2\theta_{a}=0.5\pm0.055$.
 A similar plot was presented in \cite{Miranda:2016wdr} for $L/E=500$,
 in order to understand the origin of the ambiguity in parameter space
 which is inherent to the problem.
 Now we show that, for the same baseline $L/E\approx 500$ m/MeV, 
 the uncertainties in the atmospheric mixing angle spoil the good 
 sensitivity to $\delta_{CP}$ found after the combination of neutrino and antineutrino channel in Ref.~\cite{Miranda:2016wdr}.
 Moreover, one should keep in mind that, in a realistic case, the
 existence of flux uncertainties would change each of the ellipses of
 \gfig{fig:elpseT2K} into bands.

 The reason that the leptonic Dirac CP phase $\delta_{CP}$ can be faked
 by non-unitarity at T2(H)K is due to the choice of narrow neutrino
 energy spectrum with peak around 550\,MeV and baseline at
 295\,km.
 With this choice, the oscillation phase $\Phi_a \approx \pi / 2$ is
 almost maximal and the $\cos \delta_{CP}$ term vanishes with its coefficient
 $\cos \Phi_a$. 
 It is still easy for the CP phase $\phi$ associated to non-unitarity
 to fake the standard Dirac phase $\delta_{CP}$, even at the special
 point pointed in \cite{Miranda:2016wdr}, where the degeneracies cancel
 out in the ideal case of precisely known $\theta_a$ and monochromatic
 energy spectrum.  The faking of the standard Dirac CP phase comes from
 the interplay of various elements. Around the maximal oscillation
 phase, $\Phi_a \approx \pi / 2$, the oscillation probability for
 neutrinos and anti-neutrinos can be approximated by,
 \begin{subequations}
 \begin{eqnarray}
   P_{\mu e}
 & \approx &
   4 s^2_a c^2_r s^2_r \sin^2 \Phi_a
 + 2 |\alpha_{21}| \mathbb R(S'_{11} S'^*_{13}) \cos (\phi + \delta_{CP})
 \nonumber
 \\
 & - &
   \mathbb I(S'_{12} S'^*_{13}) \sin \delta_{CP}
 +
   2 |\alpha_{21}|
   \mathbb I(S'_{11} S'^*_{12}) \sin \phi
  \,,
 \\
   P_{\bar \mu \bar e}
 & \approx &
   4 s^2_a c^2_r s^2_r \sin^2 \Phi_a
 + 2 |\alpha_{21}| \mathbb R(S'_{11} S'^*_{13}) \cos (\phi + \delta_{CP})
 \nonumber
 \\
 & + &
   \mathbb I(S'_{12} S'^*_{13}) \sin \delta_{CP}
 -
   2 |\alpha_{21}|
   \mathbb I(S'_{11} S'^*_{12}) \sin \phi
 \,,
 \end{eqnarray}
 \end{subequations}
 where the first line is the same both for neutrino and anti-neutrino
 modes, while the second receives a minus sign. To fit the current
 experimental best value $\delta^{true}_{CP} = -\pi/2$ with the
 opposite $\delta^{fit}_{CP} = \pi/2$, the major difference is
 introduced by the $\sin$ terms in the second line. The CP sensitivity
 is spoiled by freeing $\theta_a$ and $|\alpha_{21}|$ and it can be
 faked by varying $\phi$. This introduces a common correction via the
 $\cos (\phi + \delta_{CP})$ term for both neutrino and anti-neutrino
 channels. The large uncertainty in the atmospheric angle, which can
 reach $10\%$ in $s^2_a$, helps to absorb this common correction. The
 remaining $\sin \phi$ and $\sin (\phi + \delta_{CP})$ terms can then
 fake the genuine CP term $\sin \delta_{CP}$.  Although the
 coefficients of $\sin \phi$ and $\sin (\phi + \delta_{CP})$ are
 relatively small, they are not zero.  As long as $\alpha_{21}$ is
 large enough, CP can be faked. This can explain the behavior seen in
 \gfig{fig:chi2-T2K} and \gfig{fig:elpseT2K}.

 \section{Probing CP violation with $\mu$DAR and Near Detector}  
 \label{sec:muDAR}

 In order to fully resolve the degeneracy between the unitary and
 non-unitary CP phases, it is necessary to bring back the $\cos
 \delta_{CP}$ dependence by carefully choosing the energy spectrum and
 baseline configuration.  
 A perfect candidate for achieving this is to use muon decay at rest
 ($\mu$DAR) which has a wide peak and shorter baseline around 15-23
 km.  The TNT2K experiment \cite{TNT2K} is proposed to supplement the
 existing Super-K detector and the future Hyper-K detector with a
 $\mu$DAR source.  Since the accelerator neutrinos in T2(H)K have
 higher energy than those of the $\mu$DAR source, the two measurements
 can run simultaneously. Note that for T2K we use the current
   configuration as described in \gsec{sec:fake}, while for T2HK the
   $7.8 \times 10^{21} \mbox{POT}$ flux is assigned to neutrino mode
   only. On the other hand, the $\mu$DAR source can contribute a flux
   of $1.1 \times 10^{25} \mbox{POT}$ \cite{TNT2K}.  Notice that this
 experiment has backgrounds from atmospheric neutrinos, from the
 elastic scattering with electrons, and the quasi-elastic scattering
 with heavy nuclei. In addition, the $\mu$DAR flux can have 20\%
 uncertainty if there is no near detector.

\begin{figure}[t]
 \centering
 \includegraphics[width=5.6cm,angle=-90]{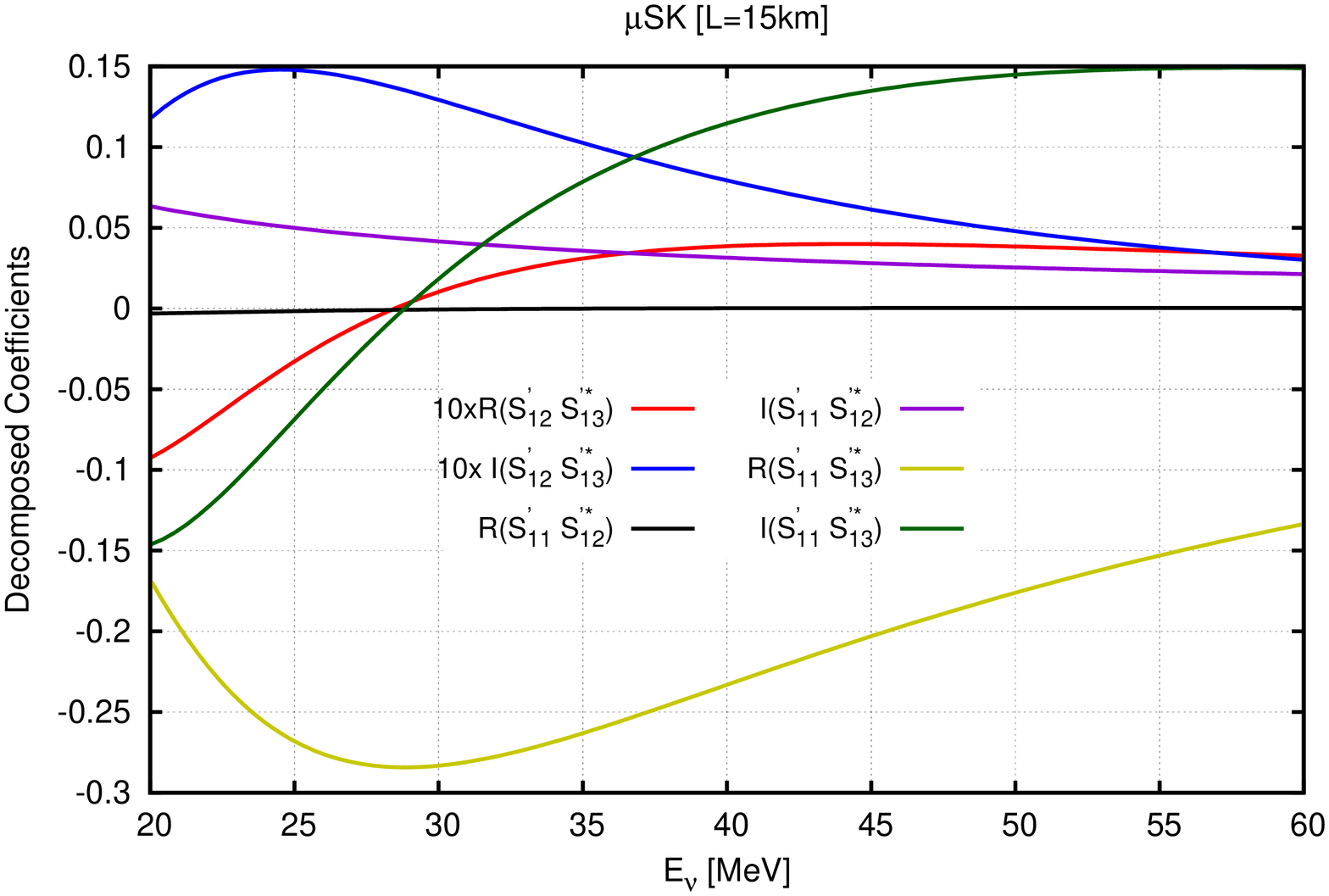}
 \includegraphics[width=5.6cm,angle=-90]{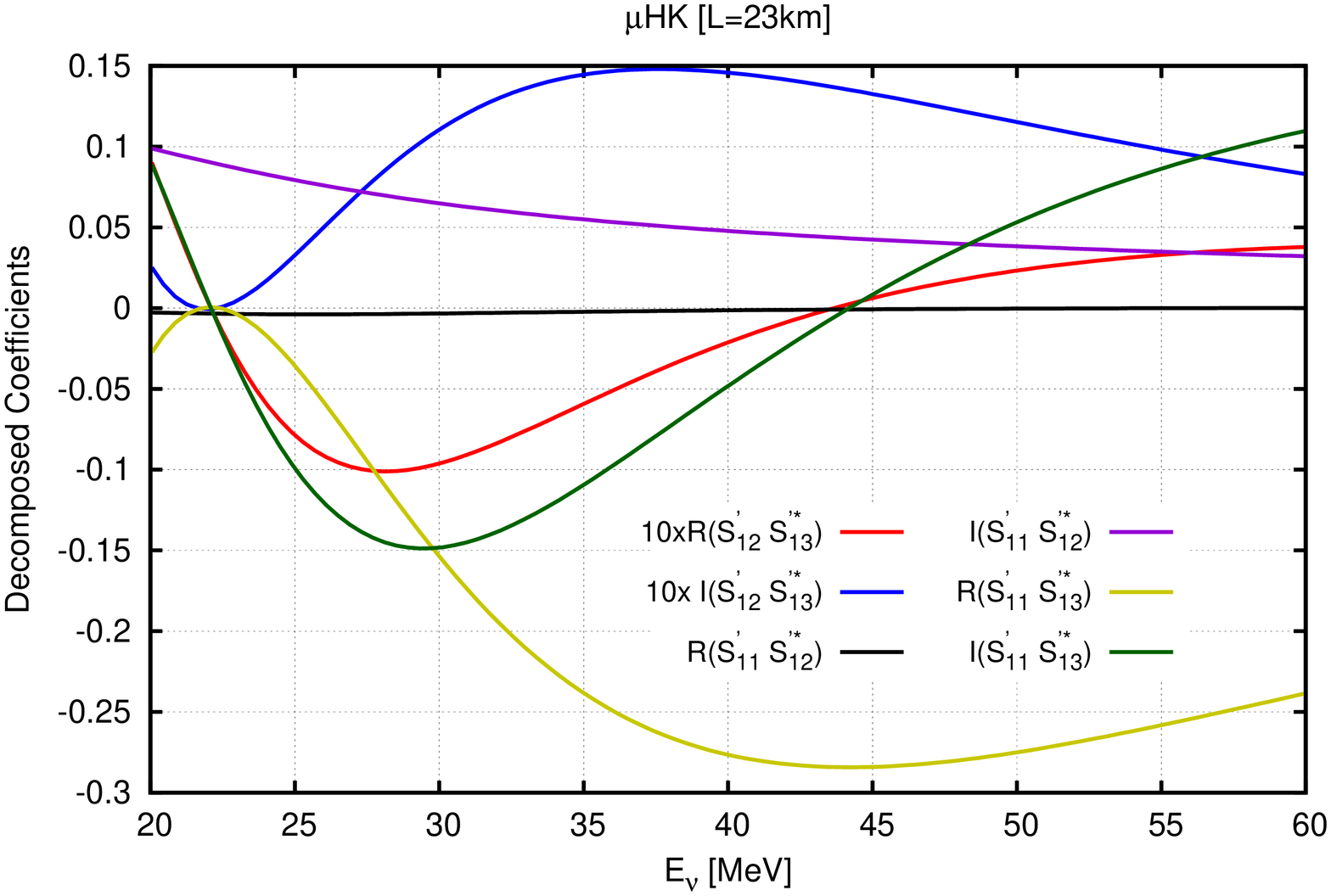}
 \caption{The amplitude matrix elements $S'_{ij}$ that contribute to
   the decomposed CP coefficients for the probabilities of
   anti-neutrino oscillation at $\mu$SK and $\mu$HK.}
 \label{fig:deCoeff-muKam}
 \end{figure}

 Note also that the sensitivity to break the degeneracy between
 $\delta_{CP}$ and $\pi - \delta_{CP}$ at T2(H)K, arising from the
 single $\sin \delta_{CP}$ dependence, can be improved because of the
 wide spectrum of $\mu$DAR, which has both $\cos \delta_{CP}$ and $\sin
 \delta_{CP}$ dependences as shown in \gfig{fig:deCoeff-muKam}.
 For the $\mu$DAR flux, the spectrum peaks around 40-50 MeV. In
 this energy range, the decomposed coefficients $P^{(2)}_{\mu e, e
   \mu}$ for the $\cos \delta_{CP}$ dependence have comparable magnitude
 with the $\sin \delta_{CP}$ term coefficients $P^{(3)}_{\mu e, e
   \mu}$.  In contrast, for T2(H)K the coefficients $P^{(2)}_{\mu e, e
   \mu}$ vanish around the spectrum peak $\sim 550$\,MeV while
 $P^{(3)}_{\mu e, e \mu}$ have sizable magnitude, as shown in
 \gfig{fig:deCoeff-T2K}. 

\begin{figure}[t]
 \centering
 \includegraphics[scale=0.31,angle=270]{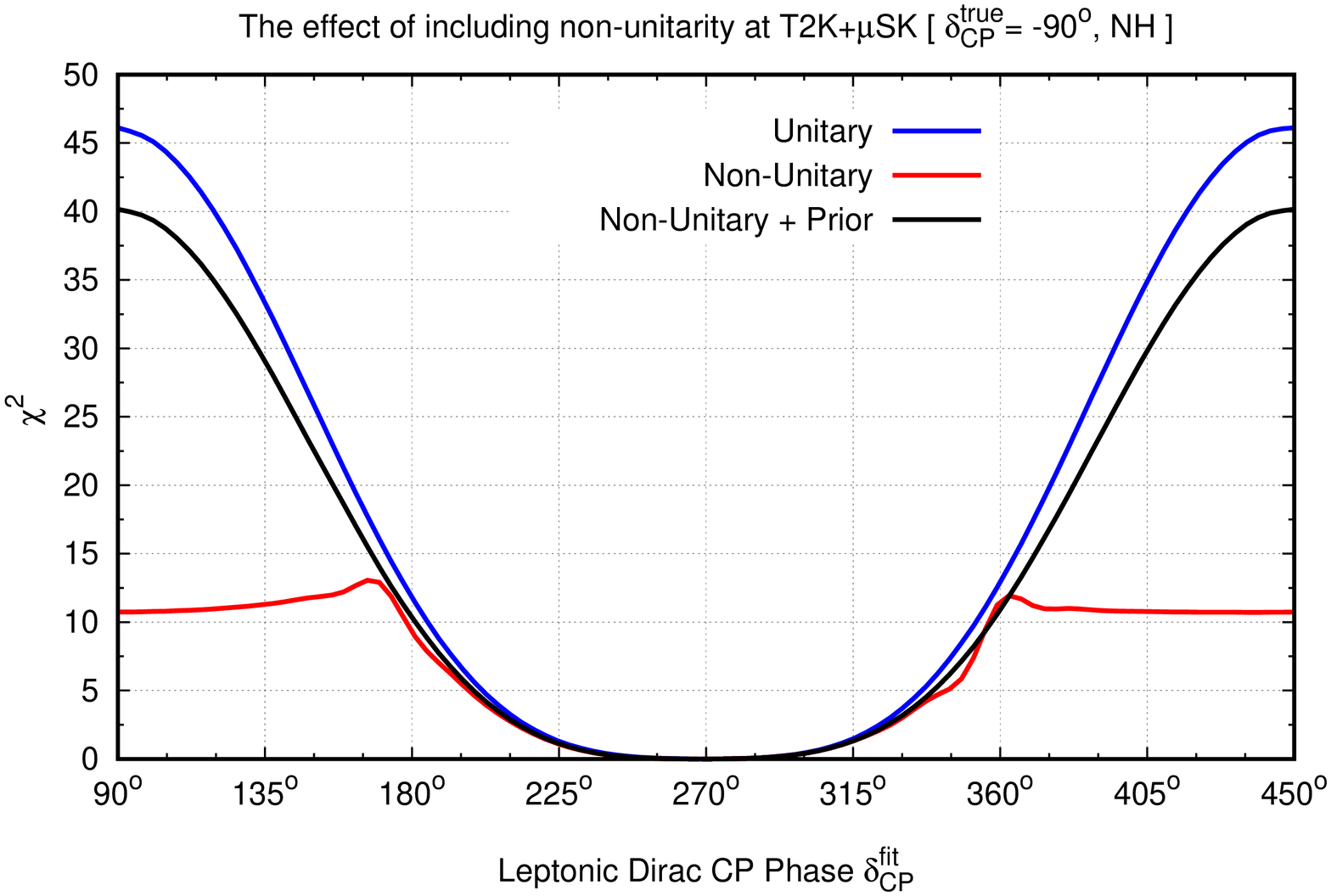}
 \includegraphics[scale=0.31,angle=270]{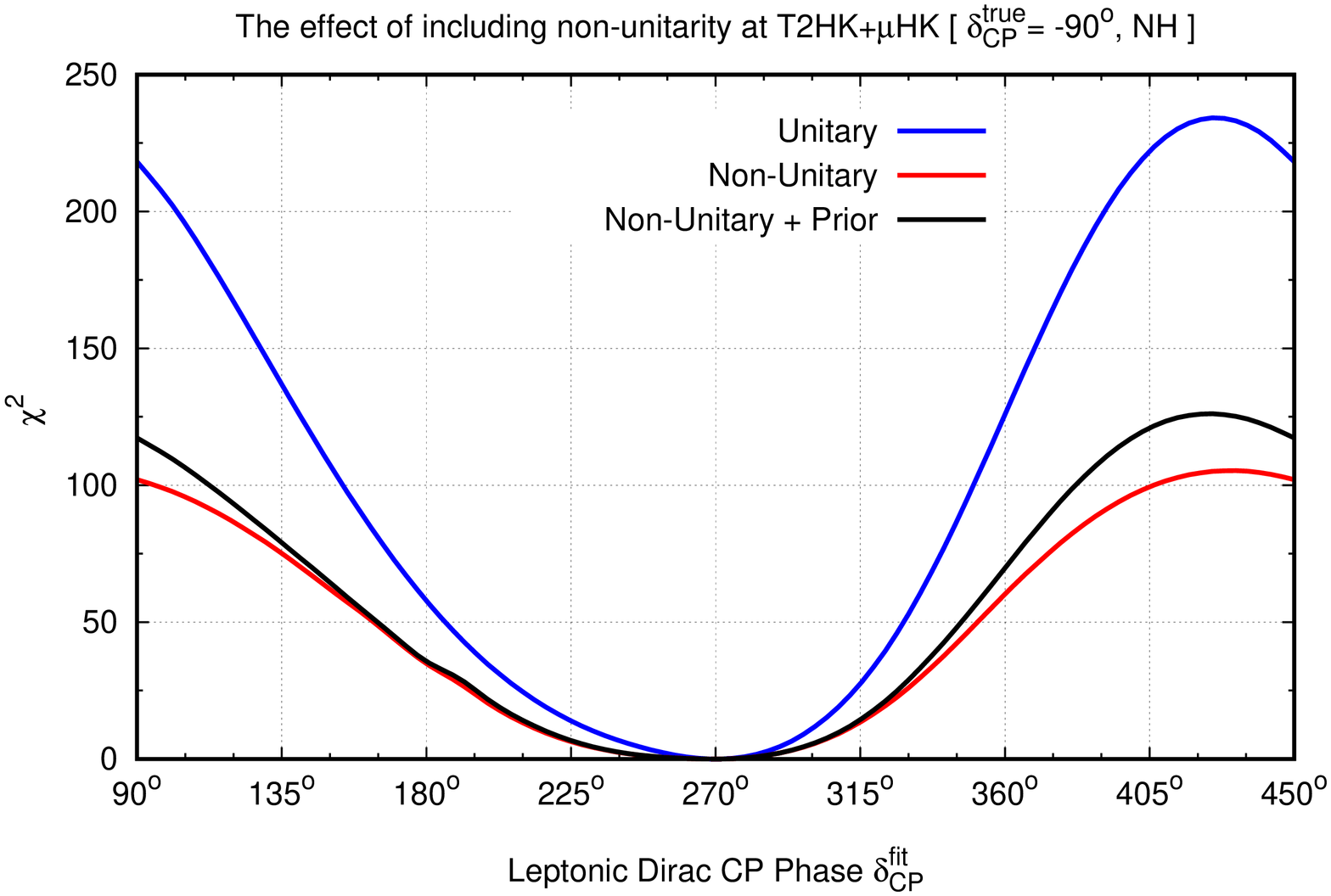}
 \caption{The marginalized $\chi^2(\delta_{CP})$ function at TNT2K
   under the assumptions of unitarity ({\color{blue}{blue}}),
   non-unitary mixing with ({\color{red}{red}}) or without (black)
   the prior constraint.}
 \label{fig:chi2-TNT2K}
 \end{figure}

 The property of having both $\cos \delta_{CP}$ and $\sin \delta_{CP}$
 dependences is exactly what we need also to break the degeneracy
 between the unitary and non-unitary CP phases. As shown in
 \gfig{fig:chi2-TNT2K}, supplementing T2K with $\mu$SK can preserve the
 CP sensitivity at the T2K level even if not imposing the prior
 constraint \geqn{eq:prior2}.  With the prior constraint, the CP
 sensitivity can further improve beyond that of T2K alone for unitary
 mixing.  The same holds for the T2HK configuration.  Nevertheless, the
 advantage of $\mu$DAR is still not fully utilized.

\begin{figure}[t!]
 \centering
 \includegraphics[scale=0.6]{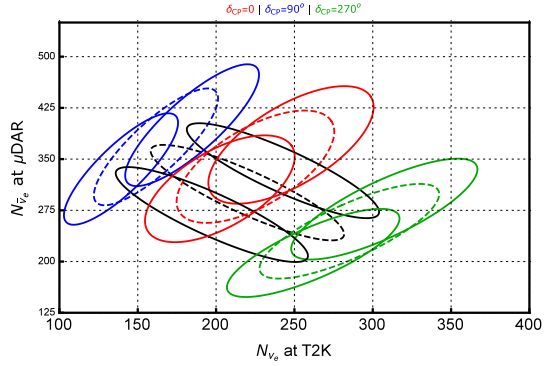}
 \caption{ Bi-event rate plot for TNT2K for standard three--neutrino
   mixing with varying $\delta_{CP}$ (black line), and non-unitary
   mixing with fixed $\delta_{CP}$ value and varying $\phi$ (color
   lines).  Dashed lines correspond to $\sin^2\theta_{a}=0.5$ while
   solid lines correspond to $\sin^2\theta_{a}=0.5\pm0.055$.  }
 \label{fig:elpseTNT2K}
 \end{figure}

 An important difference between T2(H)K in \gfig{fig:chi2-T2K} and
 TNT2K in \gfig{fig:chi2-TNT2K} is the effect of adding the prior
 constraint. At T2(H)K, the prior constraint can only add some
 moderate improvement. On the other hand, its effect can be maximized
 at TNT2K after including $\mu$Kam.  We find that the CP sensitivity
 is significantly improved by the combination of $\mu$Kam and prior
 constraints.  Notice in \gfig{fig:elpseTNT2K} that the ambiguity of
 the ellipses was not improved by having another experiment,
 nevertheless one can distinguish the standard case from the
 non-unitary case by taking a closer look at the neutrino spectrum
 which contains more information.
 Indeed, the advantage of $\mu$Kam is not fully explored with the
 current prior constraint in \geqn{eq:prior2}. Since the non-unitary CP
 effect is modulated by $|\alpha_{21}|$, a more stringent constraint on
 $|\alpha_{21}|$ would effectively suppress the size of the faked CP
 violation. From the expression of $P^{NP}_{\mu e}$ in Eq.~\geqn{eq:PNP-me},
 one sees that if the oscillation baseline is extremely short,  it is
 dominated by the last term
 \begin{equation}
   P^{NP}_{\mu e}
 \approx
   \alpha^2_{11} |\alpha_{21}|^2 \,,
 \end{equation}
 which is a nonzero constant.  Such ``zero--distance effect'' is a
 direct measure of the effective non--orthonormality of weak--basis
 neutrinos~\cite{valle:1987gv,nunokawa:1996tg}. Although $P^{NP}_{\mu
   e}$ is suppressed by $|\alpha_{21}|^2$, which is smaller than $6.6
 \times 10^{-4}$ at 90\% C.L., a near detector with a very short baseline 
 can still collect enough number of events to provide information of this parameter.

\begin{figure}[t]
 \centering
 \includegraphics[height=0.45\textwidth,width=5.5cm,angle=-90]{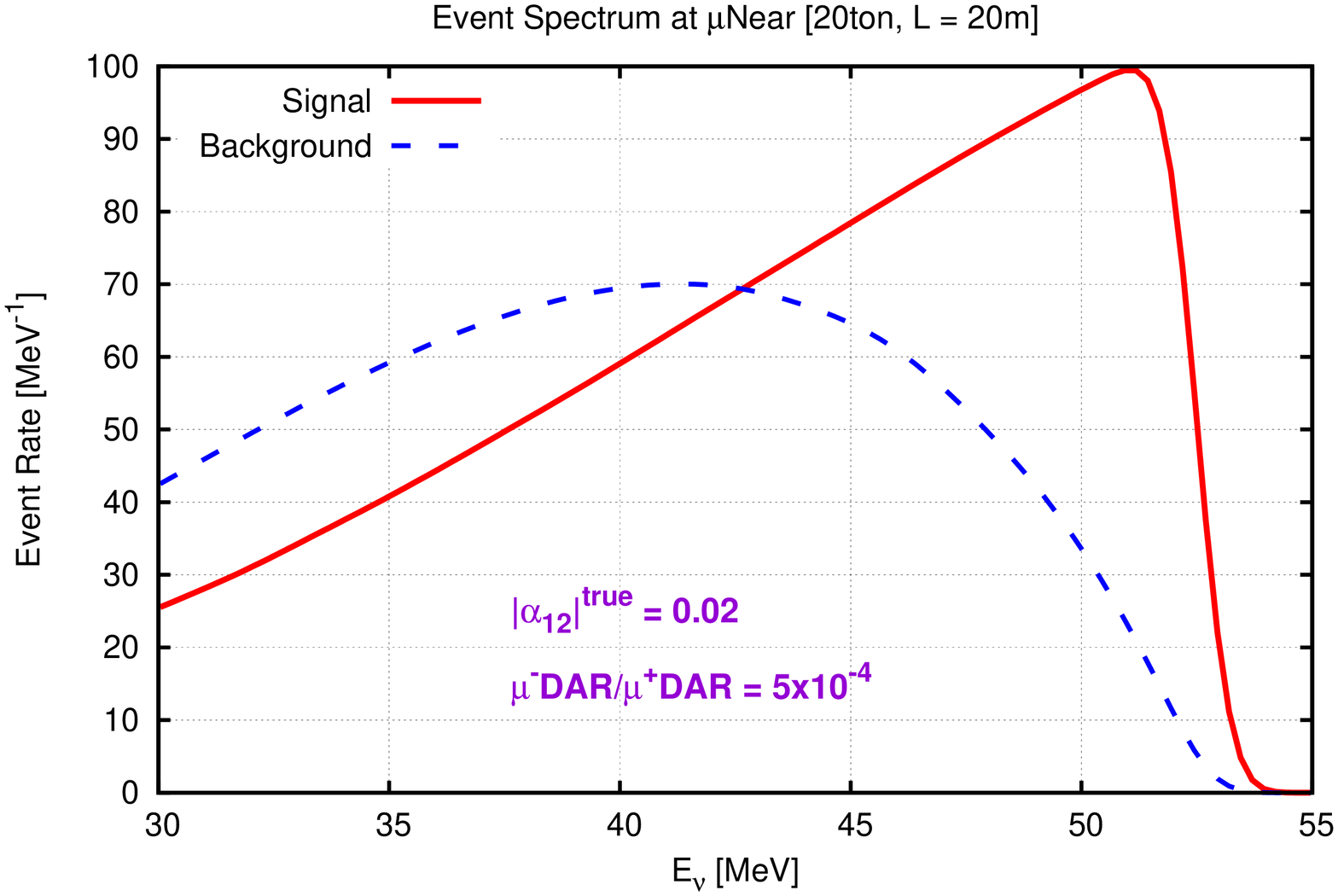}
 \includegraphics[height=0.50\textwidth,width=5.5cm,angle=-90]{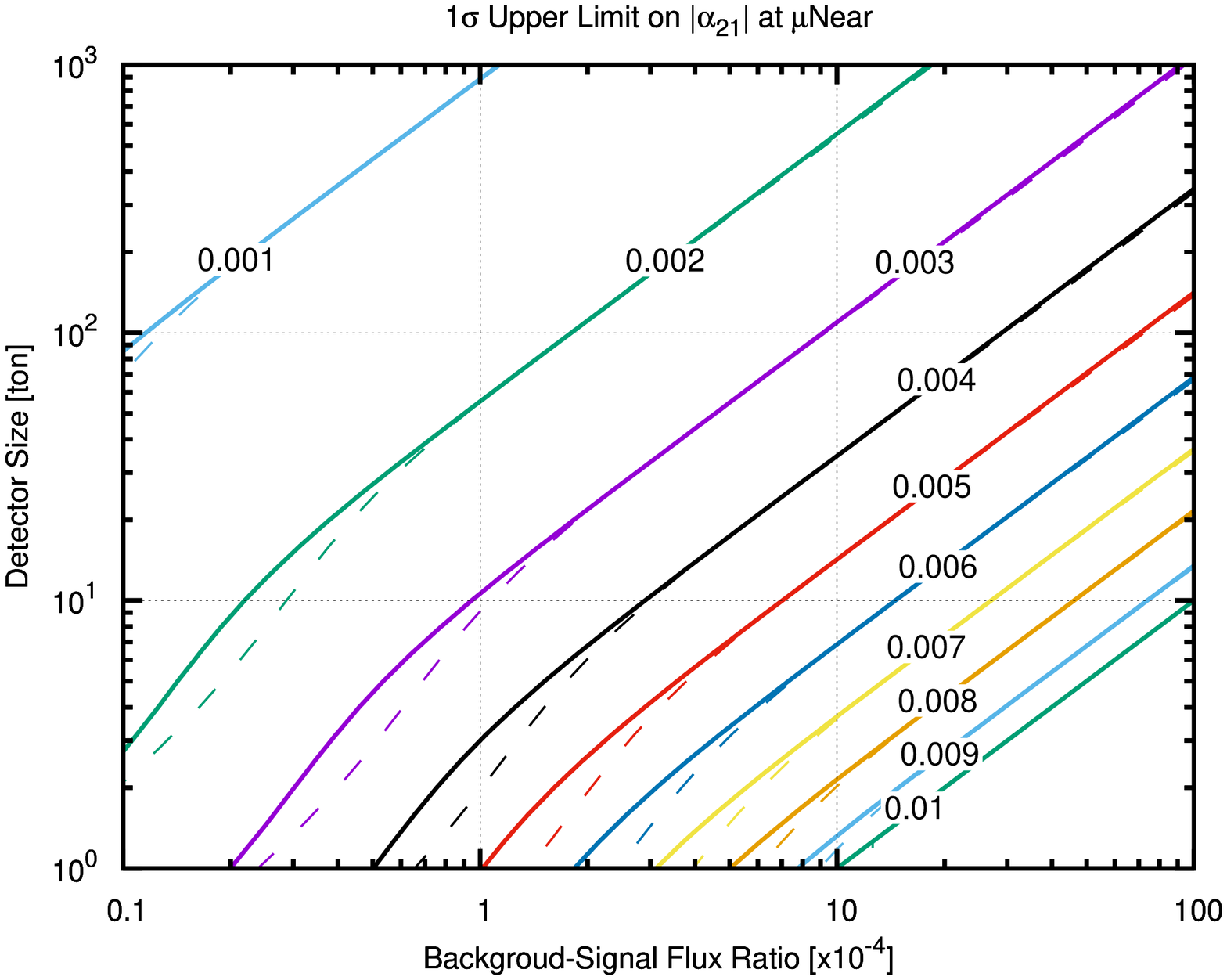}
 \caption{Event rates (left panel) and the sensitivity on
   $|\alpha_{21}|$ (right panel) at $\mu$Near as a function of
   background rate and detector size. For the sensitivity plot the
   solid contours are obtained with both 20\% uncertainty in the
   $\mu$DAR flux normalization and 50\% uncertainty in the
   background-signal flux ratio.  In contrast, the dashed contours are
   obtained with only 20\% uncertainty in the $\mu$DAR flux
   normalization while the background-signal flux ratio is fixed.}
 \label{fig:muNear}
 \end{figure}

 We propose a near detector $\mu$Near, with a 20\,ton scintillator
 detector and a 20\,m baseline to the $\mu$DAR source, to supplement the
 $\mu$Kam part of TNT2K.  By selecting events with double coincidence,
 the scintillator can identify the oscillated electron anti-neutrinos. Most of the 
 events come from two sources: the signal from $\mu^+$ decay and
 the background from $\mu^-$ decay. For both signal and background, the
 parent muons decay at rest and hence have well--defined spectrum as
 shown in the left panel of \gfig{fig:muNear}. For a background-signal
 flux ratio $\mu^-\mbox{DAR}/\mu^+\mbox{DAR} = 5 \times 10^{-4}$
 \cite{TNT2K} and non-unitary size $|\alpha_{21}| = 0.02$, the signal
 and background have roughly the same number of events, $N_{sig} =
 1446$ and $N_{bkg} = 1234$. If the neutrino mixing is unitary, only
 background is present. Based on this we can roughly estimate the
 sensitivity at $\mu$Near to be, $\sqrt{N_{bkg}}/N_{sig} \approx
 2.4\%$, for $|\alpha_{21}|^2 = (0.02)^2$. When converted to $|\alpha_{21}|$, the
 limit can be improved by a factor of $1/\sqrt{2.4\%} \approx 6.5$ on
 the basis of $0.02$ around $1\,\sigma$. In addition, the spectrum
 shape is quite different between the signal and background. The signal
 peak appears around 50\,MeV where the background event rate is much
 smaller.  This feature of different energy spectrum can further
 enhance the sensitivity than the rough estimation from total event
 rate. The constraint on $|\alpha_{21}|$ can be significantly improved
 beyond the current limit in \geqn{eq:prior2}.

 In the right panel of \gfig{fig:muNear} we show the sensitivity on
 $|\alpha_{21}|$ as a function of the background rate and the detector
 size from a simplified template fit.  The result for $5 \times 10^{-4}$
 of background and 20\,ton detector is of the same size as the rough
 estimation. The concrete value, $|\alpha_{21}| < 0.004$ at
 1\,$\sigma$, is lightly larger due to marginalization. In
 \gfig{fig:muNear} we assumed systematic errors to be 20\% for the
 $\mu$DAR flux normalization and 50\% for the background-signal flux
 ratio.  The solid contours in the right panel are obtained with both
 systematic errors imposed while the dashed ones with only the 20\%
 uncertainty in flux normalization. The difference in the sensitivity
 on $|\alpha_{21}|$ only appears in the region of small detector size
 or small background rate. For the 20\,ton detector and background rate
 larger than $10^{-4}$, the difference is negligibly small. In the full
 simulation, we only implement the 20\% uncertainty in flux
 normalization for simplicity.
 \begin{figure}[t]
 \centering
 \includegraphics[scale=0.31,angle=270]{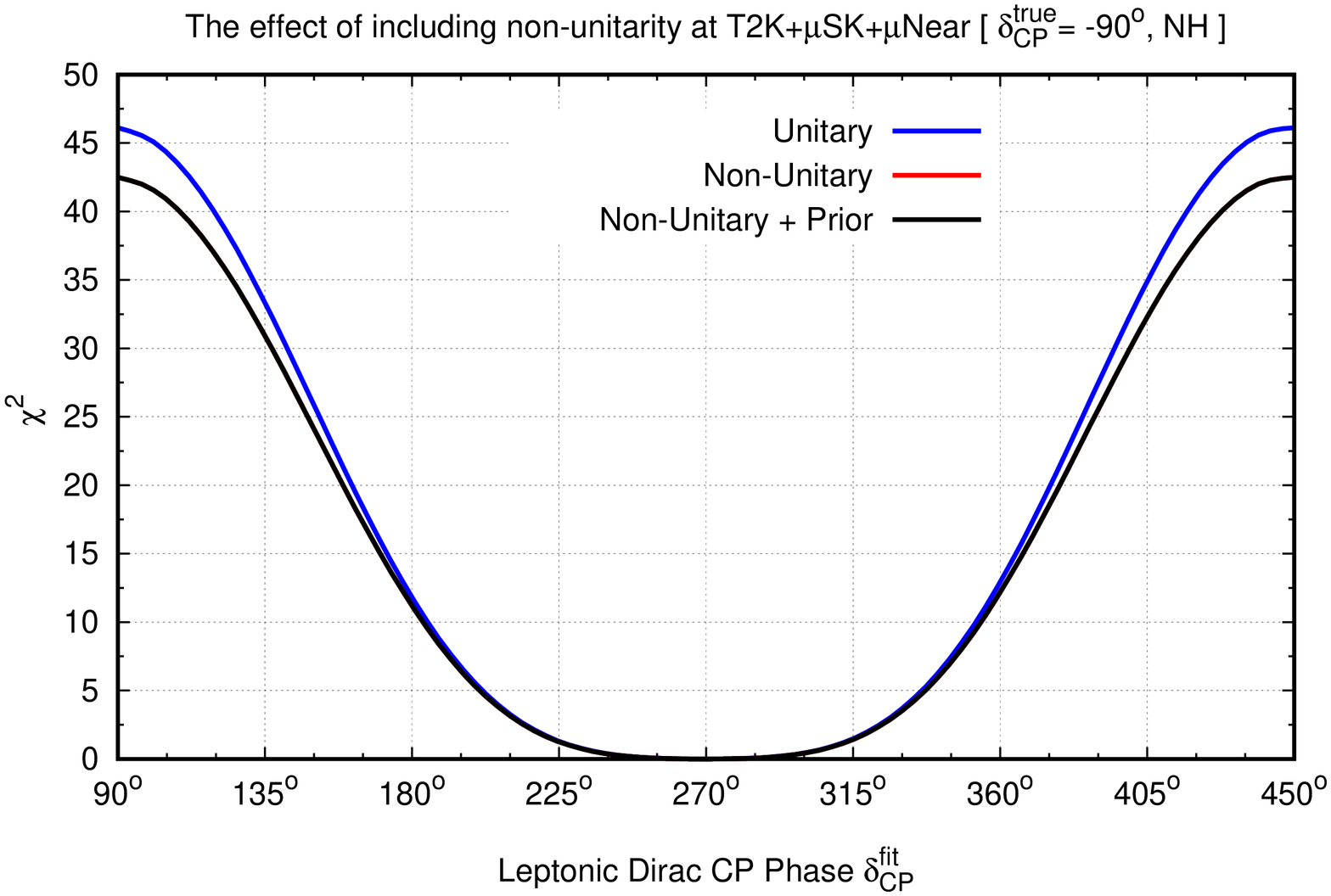}
 \includegraphics[scale=0.31,angle=270]{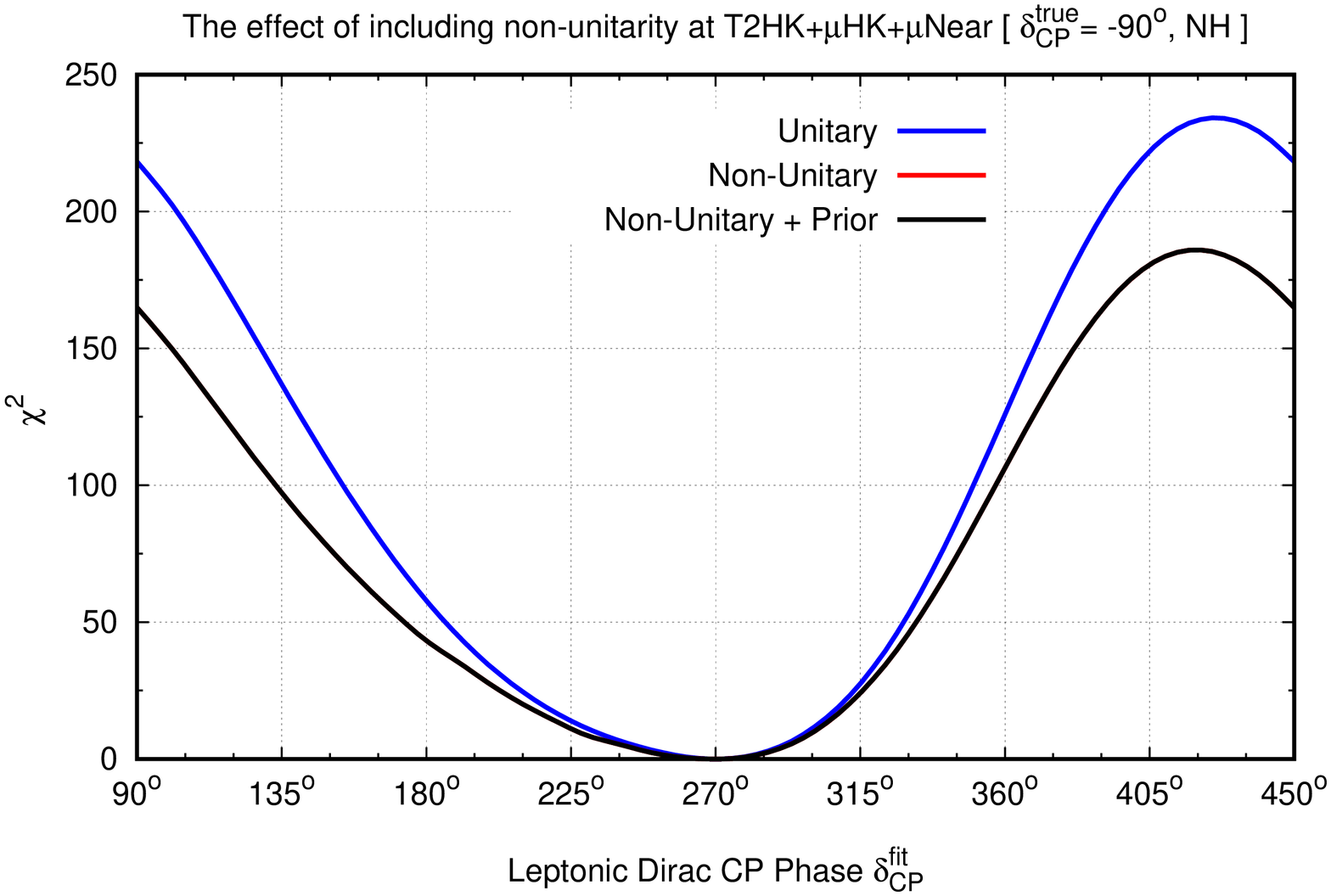}
 \caption{The marginalized $\chi^2(\delta_{CP})$ function at TNT2K +
   $\mu$Near under the assumptions of unitarity ({\color{blue}{blue}}),
   non-unitary mixing with (black) or without ({\color{red}{red}})
   the prior constraint.}
 \label{fig:chi2-muNear}
 \end{figure}
 In \gfig{fig:chi2-muNear} we show the CP sensitivity at TNT2K plus
 $\mu$Near once a full simulation is performed.  Imposing all the
 information we can get from TNT2K, $\mu$Near, and the prior
 constraint on the non-unitary mixing parameters \geqn{eq:prior2}, the
 CP sensitivity can match the full potential of TNT2K under the
 assumption of unitary mixing. Even without the prior constraint, the
 CP sensitivity at TNT2K plus $\mu$Near is very close to the full
 reach of TNT2K with unitary mixing. Imposing the prior constraint
 \geqn{eq:prior2} has little effect since the constraint on
 $\alpha_{21}$ from the $\mu$Near detector can be better by one order
 of magnitude.  This combination of CP measurements, TNT2K plus
 $\mu$Near, can determine the leptonic Dirac CP phase $\delta_{CP}$
 unambiguously and hence provide an ultimate solution to the
 degeneracy between unitary and non-unitary CP violation parameters.

 \section{Conclusion}
 \label{sec:conclusion}

 Our interpretation of experimental data always relies on theoretical
 assumptions.  Unambiguous understanding of reality always requires
 distinguishing alternative assumptions through careful experimental
 design.
 The degeneracy between unitary and non-unitary CP phases in neutrino
 mixing provides a perfect example.  In this paper we have confirmed,
 in agreement with Ref.~\cite{Miranda:2016wdr}, that, for values of
 $|\alpha_{21}|$ of the order of a few\%, one can have unitarity
 violating CP oscillation amplitudes of the same order, or possibly
 larger, than the standard one associated to $\delta_{CP}$.
 We have illustrated how the CP sensitivity at accelerator neutrino
 experiments like T2(H)K is severely degraded in the presence of
 non-unitarity. Indeed, in addition to the standard leptonic Dirac CP
 phase $\delta_{CP}$ if neutrino mixing is non-unitary there is an
 extra CP phase $\phi$ characterizing deviations from unitarity and
 affecting the neutrino appearance probability.
 The effect of such unitary phase $\delta_{CP}$ can be easily faked by
 the non-unitarity phase $\phi$ if only the $\sin \delta_{CP}$
 dependence is probed, as in the T2(H)K configuration.
 Probing the interplay with the $\cos \delta_{CP}$ dependence can help
 to lift the degeneracy.  

A perfect solution comes from the TNT2K
 project with T2(H)K supplemented by a $\mu$DAR source. Thanks to the
 different energy scale of the accelerator and $\mu$DAR neutrino
 fluxes, two different measurements can proceed at the same time, using
 Super-K and Hyper-K detectors simultaneously.
 In its original proposal, the goal was to get better measurement of
 the Dirac CP phase $\delta_{CP}$ within the standard three-neutrino
 mixing benchmark.
 We find that it also has the potential of breaking the degeneracy between
 standard and non-unitary CP phases. 
 However, TNT2K can fully explore its advantage only in combination
 with a near detector. We propose using $\mu$Near, with only 20\,ton
 of scintillator and 20\,m of baseline, to monitor the size of the
 non-unitary CP violating term for the $\mu \to e$ transition,
 $|\alpha_{21}|$. Our simplified template fit shows that $\mu$Near,
 with an expected background-signal flux ratio in the $\mu$DAR source
 of $5 \times 10^{-4}$, can constrain $|\alpha_{21}|$ to be smaller
 than $4 \times 10^{-3}$ at $1\,\sigma$, which corresponds to almost
 one order of magnitude improvement with respect to the current
 model-independent bound obtained from NOMAD data.  This estimate is
 stable against the large uncertainty in the background-signal flux
 ratio. When implemented in a full simulation, $\mu$Near can almost
 retrieve the CP sensitivity of TNT2K, providing an ultimate solution
 to the degeneracy between unitary and non-unitary mixing parameters.

 In short, non-unitary neutrino mixing is expected in a large class of
 seesaw schemes at LHC--accessible mass scales. This implies extra
 mixing parameters, and a new CP phase, that can fake the standard
 leptonic CP phase $\delta_{CP}$ present in the simplest three-neutrino
 paradigm. As a result, probing for CP violation in accelerator-type
 experiments can be misleading. We have considered T2(H)K as an example
 to illustrate the degeneracy between the ``standard'' and
 ``non-unitary'' CP phases. Despite the complete loss in its CP
 sensitivity we note that supplementing T2(H)K with a $\mu$DAR source
 can help breaking the CP degeneracy, by probing separately both $\cos
 \delta_{CP}$ and $\sin \delta_{CP}$ dependences in the wide energy
 spectrum of the $\mu$DAR flux.  We have seen that the further addition
 of a near detector to the $\mu$DAR setup has the potential of removing
 the degeneracy rather well.

 \black
 \section{Acknowledgements}

 Work supported by Spanish grants FPA2014-58183-P, Multidark
 CSD2009-00064, SEV-2014-0398 (MINECO), PROMETEOII/2014/084
 (Generalitat Valenciana).  M.~T. is supported by a Ram\'{o}n y Cajal
 contract (MINECO).  P. S. P. would like to thank the support of FAPESP
 funding grant 2014/05133-1, 2015/16809-9 and 2014/19164-6. SFG thanks Jarah Evslin
 for useful discussions.  

 \begin{appendix}

\section{Decomposition Formalism for Non-Unitary Mixing}
 \label{sec:decomposition}

 The parametrization in Eq.~\geqn{eq:N} isolates the effect of
 non-unitarity as a multiplicative matrix on the left-hand side of the
 unitary mixing matrix $U$.  This choice is extremely convenient to
 separate the neutrino oscillation probabilities into several terms,
 using the decomposition formalism \cite{Ge:2013zua}.
 The latter has a huge benefit for the case of non-unitary mixing,
 characterized by the parameters $\alpha_{ij}$ in $N^{NP}$. Indeed it
 simplifies considerably the calculation of the oscillation
 amplitudes as we demonstrate below.

  The neutrino oscillation amplitude can always be evaluated as,
\begin{equation}
  S^{n \times n}
\equiv
  e^{- i t \mathcal H^{n \times n}} \,,
\end{equation}
no matter in which basis. It is convenient to first diagonalize the
Hamiltonian,
 \begin{equation}
  \mathcal H^{n \times n}
 =
   U^{n \times n}
 \left\lgroup
 \begin{matrix}
   \sqrt{E^2 - m^2_1} \\
   & \ddots \\
   & & \sqrt{E^2 - M^2_n}
 \end{matrix}
 \right\rgroup
   (U^{n \times n})^\dagger
 \equiv
   U^{n \times n} \mathcal H^{n \times n}_D (U^{n \times n})^\dagger \,,
 \label{eq:Sp2}
 \end{equation}
 and evaluate the oscillation in the mass eigenstate basis,
\begin{equation}
  S^{n \times n}
=
   U^{n \times n} S^{n \times n}_D (U^{n \times n})^\dagger \,.
\end{equation}
 For neutrino
 oscillation at low energy, $E < M_{4, \cdots, n}$, the heavy state decays
 with an imaginary Hamiltonian. In other words, the oscillation amplitude
 matrix $S^{n \times n}_D \equiv e^{- i t \mathcal H^{n \times n}_D}$ in
 the mass eigenstate basis has non-trivial elements only in the $3 \times 3$
 light block. The oscillation within the three light neutrinos can then
 be described by the effective amplitude matrix,
\begin{equation}
   S^{NP}
 =
   N^{NP} S N^{NP \dagger} \,,
 \label{eq:SNP}
 \end{equation}
 where $S$ is the standard amplitude matrix corresponding to unitary
 mixing $U$.
 Note that the extra neutrinos are much heavier than the energy scale
 under discussion and hence decouple from the (low-energy) neutrino
 oscillations.  Their low-energy effect
 is just a basis transformation which also applies to the oscillation
 amplitudes. 
 The neutrino oscillation probability is given by the squared magnitude
 of the corresponding amplitude matrix element, $P^{NP}_{\alpha \beta}
 = |S^{NP}_{\beta \alpha}|^2$,
 \begin{subequations}\label{eq:probtotal}
 \begin{eqnarray}
   P^{NP}_{ee}
 & = &
   \alpha^4_{11} P_{ee} \,,
 \\
   P^{NP}_{e \mu}
 & = &
   \alpha^2_{11}
 \left[
   \alpha^2_{22} P_{e \mu}
 + 2 \alpha_{22} \mbox{Re} \left( \alpha_{21} S_{ee}^* S_{\mu e} \right)
 + |\alpha_{21}|^2 P_{ee}
 \right] \,,
 \\
   P^{NP}_{\mu e}
 & = &
   \alpha^2_{11}
 \left[
   \alpha^2_{22} P_{\mu e}
 + 2 \alpha_{22} \mbox{Re} \left( \alpha_{21}^* S_{ee} S_{e \mu }^* \right)
 + |\alpha_{21}|^2 P_{ee}
 \right] \,,
 \\\nonumber
   P^{NP}_{\mu \mu}
 & = &
   \alpha_{22}^4 P_{\mu\mu}
 + |\alpha_{21}^2| \alpha_{22}^2 (P_{\mu e} + P_{e \mu})
 + |\alpha_{21}|^4 P_{ee}
 \\  
 & + &
   \sum_{\{a_1,b_1\}\neq\{a_2,b_2\}}\rm{Re}[\alpha_{2a_1}\alpha_{2b_1}^*\alpha_{2a_2}^*\alpha_{2b_2}S_{a_1b_1}S_{a_2b_2}^*].
 \end{eqnarray}
 \label{eq:prob-nonunitary}
 \end{subequations}
 Here $P_{\alpha \beta}$ is the oscillation probability with unitary
 mixing and ($a,b$)=($1,2$) for $\alpha_{ab}$ while ($a,b$)=($e,\mu$)
 for $S_{ab}$.
 Note that the remaining five oscillation probabilities ($P^{NP}_{e
   \tau}$, $P^{NP}_{\tau e}$, $P^{NP}_{\mu \tau}$, $P^{NP}_{\tau \mu}$,
 $P^{NP}_{\tau \tau}$) can not be derived from the four in
 \geqn{eq:prob-nonunitary} by unitarity conditions since these do not
 hold in our case. Instead, they need to be calculated directly from
 $S^{NP}$ elements in a similar way as the above four.

 In addition, the atmospheric mixing angle and the Dirac CP phase
 $\delta_{CP}$ can also be factorized out as transformations,
 \begin{equation}
   \mathcal H
 =
   [U_{23}(\theta_a) P_\delta] \mathcal H' [U_{23}(\theta_a) P_\delta]^\dagger \,,
 \qquad
   S
 =
   [U_{23}(\theta_a) P_\delta] S' [U_{23}(\theta_a) P_\delta]^\dagger \,,
 \label{eq:HS}
 \end{equation}
 where $U_{23}(\theta_a)$ is the 2--3 mixing parameter and $P_\delta
 \equiv \mbox{diag}(1,1,e^{i \delta_{CP}})$ is a rephasing matrix.
 Those quantities with prime, $\mathcal H'$ and $S'$, are defined in
 the so-called ``propagation basis''
 \cite{Akhmedov:1998xq,Yokomakura:2002av}. The connection between the
 non-unitary flavor basis and the ``propagation basis'' is $N^{NP}
 U_{23}(\theta_a) P_\delta$
 Replacing the unitary oscillation amplitude $S$ in the flavor basis by
 $S'$ \cite{Ge:2013zua} in the ``propagation basis'' with $\theta_a$
 and $\delta_{CP}$ rotated away, the non-unitary oscillation
 probabilities \geqn{eq:prob-nonunitary} become,
 \begin{subequations}
 \begin{eqnarray}
   P^{NP}_{ee}
 & = &
   \alpha^4_{11} P_{ee} \,,
 \\
   P^{NP}_{e \mu}
 & = &
   \alpha^2_{11}
 \left\{
   \alpha^2_{22} P_{e \mu}
 + 2 \alpha_{22} |\alpha_{21}|
 \left[
   c_a \left( c_\phi \mathbb R + s_\phi \mathbb I \right) (S'_{11} S'^*_{21})
 \right.
 \right.
 \nonumber
 \\
 &&
 \left.
 \left.
 \hspace{40mm}
 + s_a \left( c_{\phi + \delta_{CP}} \mathbb R + s_{\phi + \delta_{CP}} \mathbb I \right) (S'_{11} S'^*_{31})
 \right]
 + |\alpha_{21}|^2 P_{ee}
 \right\} \,,
 \\
   P^{NP}_{\mu e}
 & = &
   \alpha^2_{11}
 \left\{
   \alpha^2_{22} P_{\mu e}
 + 2 \alpha_{22} |\alpha_{21}|
 \left[
   c_a \left( c_\phi \mathbb R - s_\phi \mathbb I \right) (S'_{11} S'^*_{12})
 \right.
 \right.
 \nonumber
 \\
 &&
 \left.
 \left.
 \hspace{40mm}
 + s_a \left( c_{\phi + \delta_{CP}} \mathbb R - s_{\phi + \delta_{CP}} \mathbb I \right) (S'_{11} S'^*_{13})
 \right]
 + |\alpha_{21}|^2 P_{ee}
 \right\} \,,
 \label{eq:PNP-me}
 \\
   P^{NP}_{\mu \mu}
 & = &
 \left|
   \alpha^2_{22} S_{\mu \mu}
 + \alpha_{22} \left( \alpha_{21} S_{e\mu} + \alpha^*_{21} S_{\mu e}  \right)
 + \alpha_{11}^2 S_{ee}
 \right|^2 \,
 \end{eqnarray}
 \label{eq:PNP-expanded}
 \end{subequations}
 For convenience, we have denoted $(c_\phi, s_\phi) \equiv (\cos \phi,
 \sin \phi)$ and $(c_{\phi + \delta_{CP}}, s_{\phi +\delta_{CP}})
 \equiv (\cos(\phi + \delta_{CP}), \sin(\phi + \delta_{CP}))$, where
 $\delta_{CP}$ and $\phi$ are the leptonic Dirac CP phase and the
 non-unitary phase associated with $\alpha_{21} \equiv |\alpha_{21}|
 e^{-i \phi}$, respectively. The real and imaginary operators, $\mathbb
 R$ and $\mathbb I$, extract the corresponding part of the following
 terms. The general expression \geqn{eq:PNP-expanded} reproduces the
 fully expanded form in \cite{Escrihuela:2015wra} up to the leading
 order of $\sin \theta_r \sim 0.15$ and $\Delta m^2_s / \Delta m^2_a
 \sim 3\%$.

 The oscillation probabilities $P^{NP}_{e \mu}$ and $P^{NP}_{\mu e}$ in
 \geqn{eq:PNP-expanded} are not just functions of their unitary
 counterparts $P_{e \mu}$ and $P_{\mu e}$, but they also contain
 non-unitary CP terms involving $\phi$.
 Therefore, the non-unitarity of the neutrino mixing matrix
 introduces extra decomposition coefficients in addition to those
 proposed in \cite{Ge:2013zua},
 \begin{eqnarray}
   P^{NP}_{\alpha \beta}
 & \equiv &
   P^{(0)}_{\alpha \beta}
 + P^{(1)}_{\alpha \beta} x_{\rm a}
 + P^{(2)}_{\alpha \beta} \cos \delta'_{CP}
 + P^{(3)}_{\alpha \beta} \sin \delta'_{CP}
 + P^{(4)}_{\alpha \beta} x_{\rm a} \cos \delta'_{CP}
 + P^{(5)}_{\alpha \beta} x^2_{\rm a}
 + P^{(6)}_{\alpha \beta} \cos^2 \delta'_{CP} \qquad
 \nonumber
 \\
 & + &
   P^{(7)}_{\alpha \beta} c_a c_\phi
 + P^{(8)}_{\alpha \beta} c_a s_\phi
 + P^{(9)}_{\alpha \beta} s_a c_{\phi + \delta_{CP}}
 + P^{(10)}_{\alpha \beta} s_a s_{\phi + \delta_{CP}} \,.
 \end{eqnarray}
 Here, we have expanded the atmospheric angle $\theta_a$ around its
 maximal value $c^2_a = (1+x_a)/2$ and rescaled Dirac CP functions
 $(\cos \delta'_{CP}, \sin \delta'_{CP}) \equiv 2 c_a s_a (\cos \delta_{CP},
 \sin \delta_{CP})$.  The explicit form of these decomposition
 coefficients are shown in \gtab{tab:Ps}.
 \begin{table}[t]
 \centering
 \begin{tabular}{c|cccc}
  & $P^{(k)}_{\rm ee}$ & $P^{(k)}_{\rm e\mu}$ & $P^{(k)}_{\mu \rm e}$ \\[1mm]
 \hline
 \hline
 (0) & $\alpha^4_{11} |S'_{11}|^2$ 
     & $\alpha^2_{11} \left[ \frac {\alpha^2_{22}} 2 (1 - |S'_{11}|^2) + |\alpha_{21}|^2 |S'_{11}|^2 \right]$ 
     & $\alpha^2_{11} \left[ \frac {\alpha^2_{22}} 2 (1 - |S'_{11}|^2) + |\alpha_{21}|^2 |S'_{11}|^2 \right]$ \\[1mm]
 (1) & 0 & $\frac {\alpha^2_{11} \alpha^2_{22}} 2 (|S'_{21}|^2 - |S'_{31}|^2)$ & $\frac {\alpha^2_{11} \alpha^2_{22}} 2 (|S'_{12}|^2 - |S'_{13}|^2)$ \\[1mm]
 (2) & 0 & $\alpha^2_{11} \alpha^2_{22}\mathbb R(S'_{21} S'^*_{31})$ & $\alpha^2_{11} \alpha^2_{22}\mathbb R(S'_{12} S'^*_{13})$ \\[1mm]
 (3) & 0 & $\alpha^2_{11} \alpha^2_{22}\mathbb I(S'_{21} S'^*_{31})$ & $-\alpha^2_{11} \alpha^2_{22}\mathbb I(S'_{12} S'^*_{13})$ \\[1mm]
 (4) & 0 & 0 & 0 \\[1mm]
 (5) & 0 & 0 & 0 \\[1mm]
 (6) & 0 & 0 & 0 \\
 \hline
 (7)  & 0 & $+ 2 \alpha^2_{11} \alpha_{22} |\alpha_{21}| \mathbb R (S'_{11} S'^*_{21})$
          & $+ 2 \alpha^2_{11} \alpha_{22} |\alpha_{21}| \mathbb R (S'_{11} S'^*_{12})$ \\[1mm]
 (8)  & 0 & $+ 2 \alpha^2_{11} \alpha_{22} |\alpha_{21}| \mathbb I (S'_{11} S'^*_{21})$
          & $- 2 \alpha^2_{11} \alpha_{22} |\alpha_{21}| \mathbb I (S'_{11} S'^*_{12})$ \\[1mm]
 (9)  & 0 & $+ 2 \alpha^2_{11} \alpha_{22} |\alpha_{21}| \mathbb R (S'_{11} S'^*_{31})$
          & $+ 2 \alpha^2_{11} \alpha_{22} |\alpha_{21}| \mathbb R (S'_{11} S'^*_{13})$ \\[1mm]
 (10) & 0 & $+ 2 \alpha^2_{11} \alpha_{22} |\alpha_{21}| \mathbb I (S'_{11} S'^*_{31})$
          & $- 2 \alpha^2_{11} \alpha_{22} |\alpha_{21}| \mathbb I (S'_{11} S'^*_{13})$
 \end{tabular}
 \caption{The decomposed coefficients $P^{(k)}_{ee}$, $P^{(k)}_{e \mu}$, and
   $P^{(k)}_{\mu e}$ as an extension
   to the results first derived in \cite{Ge:2013zua}. For symmetric
   matter potential profile, the amplitude matrix $S'$ is
   also symmetric.}
 \label{tab:Ps}
 \end{table}
 For simplicity, we show just the three channels ($P^{NP}_{ee}$,
 $P^{NP}_{e \mu}$ and $P^{NP}_{\mu e}$) in \gtab{tab:Ps} to illustrate
 the idea.  Ignoring matter effects (or if these can be approximated by
 a symmetric/constant potential), the amplitude matrix $S'$ is then symmetric, 
 $S'_{ij}=S'_{ji}$. To obtain the
 anti-neutrino coefficients $\overline P^{NP}_{\alpha \beta}$, the CP
 phases ($\delta_{CP}$ and $\phi$) as well as the matter potential
 inside the $S'$ matrix elements should receive a minus sign.

 \section{Matter effect with non-unitary mixing}
 \label{sec:matter}
 The decomposition formalism presented in \gapp{sec:decomposition} is
 a powerful tool to obtain a complete formalism for neutrino
 oscillations. It factorizes the mixings efficiently in different
 bases and treats their effects independently.  For example, the
 matter potential does not spoil the relations \geqn{eq:probtotal}
 that follow from the general parametrization \geqn{eq:N}. Although
 the previous results are obtained for vacuum oscillations, one can
 still use \geqn{eq:probtotal} for neutrino oscillation through
 matter, as long as $S_{ij}$ is replaced by the corresponding
 amplitude matrix in matter, $S^{matter}_{ij}$. In this appendix we
 will show how the presence of non-unitary neutrino mixing results in
 a rescaling of the standard matter potential. Our result applies
 generally for any number of heavy neutrinos \footnote{An expansion in
   the mass hierarchy parameter $\alpha \equiv \Delta m^2_s / \Delta
   m^2_a$ and the unitarity violation parameters up to first order can
   also be found in \cite{Li:2015oal}, where they are denoted as
   $s^2_{ij}$, for $i = 1,2,3$ and $j = 4,5,6$.}.

   In order to further develop the formalism established in
   \gapp{sec:decomposition} to introduce matter effects with
   non-unitary mixing, it is extremely useful to use the symmetrical
   parametrization method for unitary matrices.
   We start by recalling that its main ingredient consists in
   decomposing $U^{n \times n}$ in terms of products of effectively
   two--dimensional complex rotation matrices $\omega_{1j}$, in which
   each factor is characterized by both one rotation angle and one CP
   phase, see Eqs.(3.9)--(3.15) and (3.19)--(3.22)
   in~\cite{Schechter:1980gr}. The method is equivalent to the
   procedure of obtaining the current PDG form of the lepton mixing
   matrix and any generalization thereof.
 In the presence of \SM singlet neutrinos, it can be used to describe
 the mixing matrix $U^{n \times n}$ as follows 
 \begin{equation}
   U^{n \times n}
 =
   \left( \Pi^n_{i > j > 3} \omega_{ij} \right)
   \left( \Pi^n_{j = 4} \omega_{3j} \right)
   \left( \Pi^n_{j = 4} \omega_{2j} \right)
   \left( \Pi^n_{j = 4} \omega_{1j} \right)
 \left\lgroup
 \begin{matrix}
   \omega_{23} P_\delta \omega_{13} \omega_{12} & 0 \\
   0 & 1
 \end{matrix}
 \right\rgroup \,,
 \end{equation}
 in the same way as for its $3 \times 3$ counterpart $U$. 
With such parametrization for the extended mixing
 matrix, one can still resort to the ``propagation basis''.
This can be achieved by dividing the full mixing matrix
$U^{n \times n} \equiv \mathcal R' U'$,
 \begin{equation}
   \mathcal R'
 =
   U^{NP}
 \left\lgroup
 \begin{matrix}
   \omega_{23} P_\delta & 0 \\
   0 & 1
 \end{matrix}
 \right\rgroup \,,
\qquad
  U'
=
 \left\lgroup
 \begin{matrix}
   \omega_{13} \omega_{12} & 0 \\
   0 & 1
 \end{matrix}
 \right\rgroup \,.
 \label{eq:Tp}
 \end{equation}
 The ``propagation basis'' is connected to
 the non-unitary flavor basis with the transformation matrix $\mathcal R'$
 and the remaining mixing is $U'$.

 The original $n\times n$  Hamiltonian is given by
 \begin{equation}
  \mathcal H^{n \times n} \!\!
 =
   U^{n \times n} \!\!\!
 \left\lgroup
 \begin{matrix}
   \sqrt{E^2 - m^2_1} \\
   & \ddots \\
   & & \sqrt{E^2 - M^2_n}
 \end{matrix}
 \right\rgroup \!\!\!
   (U^{n \times n})^\dagger \!
 + \!\!
 \left\lgroup
 \begin{matrix}
   V_{cc} \\
   & 0 \\
   & & 0 \\
   & & & \ddots
 \end{matrix}
 \right\rgroup \!\!
 + \!
   V_{nc} \!\!
 \left\lgroup
 \begin{matrix}
   1 \\
   & 1 \\
   & & 1 \\
   & & & \ddots
 \end{matrix}
 \right\rgroup
 \,,
 \label{eq:Sp2}
 \end{equation}
 We denote the matter potential matrices as $\mathbb V \equiv \mathbb V_{cc} + \mathbb V_{nc}$
 in latter discussions. For heavy mass eigenstates with $M_i > M_Z
 \gg E$, the oscillation will decay out very quickly since the
 oscillation phase $\sqrt{E^2 - M^2_n}$ is imaginary.
 For convenience, we separate the matrices into light and heavy blocks,
 \begin{equation}
  \mathcal H^{n \times n}
 =
  \mathcal R'
 \left[ U'
 \left\lgroup
 \begin{matrix}
   \sqrt{E^2 - \mathbb M^2_l} \\
 & \sqrt{E^2 - \mathbb D^2_h}
 \end{matrix}
 \right\rgroup U'^\dagger
 +
   \mathcal R'^\dagger
 \left\lgroup
 \begin{matrix}
   \mathbb V \\
   & 0
 \end{matrix}
 \right\rgroup
   \mathcal R'
 \right]
   \mathcal R'^\dagger
 \label{eq:Sp3}
 \end{equation}
 where $\sqrt{E^2 - \mathbb M^2_l}$ is the standard momentum matrix in
 the ``propagation basis'', with the solar and reactor angles
 $\theta_s$ and $\theta_r$ incorporated, while $\sqrt{E^2 - \mathbb
   D^2_h}$ is already diagonal.  As long as $\mathbb V \ll \sqrt{E^2 -
   \mathbb D^2_h}$, the mixing between the light and heavy blocks
 inside the bracket is highly suppressed by a factor of $\mathbb V /
 \sqrt{E^2 - \mathbb D^2_h}$.  For CP measurement experiments, $\mathbb
 V \lesssim \Delta m^2_a/2 E \ll \sqrt{E^2 - \mathbb D^2_h}$ with
 $\Delta m^2_a \sim \mathcal O(0.01 \, \mbox{eV}^2)$, $10\,\mbox{MeV}
 \lesssim E \lesssim 1\,\mbox{GeV}$, and $\mathbb D^2_h > M_Z^2 $,
 the induced mixing $\mathbb V/\sqrt{E^2 - \mathbb D^2_h} \lesssim
 10^{-19}$ is negligibly small.  In addition, the mixing term is
 further suppressed by the small non-unitary mixing contained in
 $\mathcal R'$.  As a good approximation for low-energy neutrino
 oscillation experiment, the light and heavy blocks decouple from each
 other.  
 We have showed that the ``propagation basis''
 \cite{Akhmedov:1998xq,Yokomakura:2002av} can still be established in the presence of
 non-unitary mixing.  Note that $\mathcal R'$ is exactly $N^{NP}
 U_{23}(\theta_a) P_\delta$ that already used in \gapp{sec:decomposition} to
 relate the non-unitary flavor basis and the ``propagation basis''
 through \geqn{eq:SNP} and \geqn{eq:HS}. In other words, as long as the
 mass of heavy neutrino is much larger than the oscillation energy and
 matter effect, the same ``propagation basis'' can be generalized for
 non-unitary mixing.

 Since the light and heavy blocks effectively decouple from each other, the
 oscillation probability can be evaluated independently. For the light block,
 we can first evaluate the amplitude matrix $S' = e^{- i \mathcal H' t}$ in the ``propagation basis''
 and transform back to the flavor basis with $\mathcal R'$ in the same way as
 \geqn{eq:PNP-expanded}. The only change is a modified matter potential,
 \begin{equation}
   \widetilde{\mathbb M}^2_l
 =
   \mathbb M^2_l
 -
   2 E \mathbb R'^\dagger \mathbb V \mathbb R' \,,
 \label{eq:M2l}
 \end{equation}
 where $\mathbb R'$ is the light block of $\mathcal R'$. Here we have
 expanded the neutrino momentum of light neutrinos in relativistic limit.  The potential
 matrix in the ``propagation basis'' is replaced by $\mathbb V
 \rightarrow \mathbb R'^\dagger \mathbb V \mathbb R'$.

 \end{appendix}



  \providecommand{\url}[1]{\texttt{#1}}
  \providecommand{\urlprefix}{URL }
  \providecommand{\eprint}[2][]{\url{#2}}


%

\end{document}